\documentclass[%
 aip,
 preprint,%
]{revtex4-1}
\usepackage[utf8]{inputenc}
\usepackage[T1]{fontenc}
\usepackage{graphicx}
\usepackage{dcolumn}
\usepackage{bm}
\usepackage{mathptmx}
\usepackage{amsmath}
\usepackage{amssymb}
\usepackage{color}
\usepackage{url}
\usepackage{nameref}
\usepackage{hyperref}

\newcommand{\dee}{\partial}
\newcommand{\flux}{\mathrm{Flux}}
\newcommand{\jsource}{J_{01}}
\newcommand{\jsink}{J_{89}}
\newcommand{\jij}{J_{45}}
\newcommand{\jim}{J_{i-1,i}}
\newcommand{\jip}{J_{i,i+1}}
\newcommand{\jjk}{J_{56}}
\newcommand{\jkl}{J_{67}}
\newcommand{\jss}{J_{ss}}

\newcommand{\kij}{k_{ij}}
\newcommand{\kim}{k_{i+1,i}}
\newcommand{\kip}{k_{i,i+1}}
\newcommand{\kji}{k_{ji}}
\newcommand{\mfpt}{\mathrm{MFPT}}
\newcommand{\psource}{P_0}
\newcommand{\tmax}{t_{\mathrm{max}}}
\newcommand{\xsrc}{x_{\mathrm{src}}}
\newcommand{\ymax}{y_{\mathrm{max}}}

\begin{document}

\preprint{AIP/123-QED}

\title{Transient probability currents provide upper and lower bounds on non-equilibrium steady-state currents in the Smoluchowski picture}




\author{Jeremy Copperman}
\affiliation{Department of Biomedical Engineering, Oregon Health and Science University, Portland, OR}
\author{David Aristoff}
\affiliation{Department of Mathematics, Colorado State University, Fort Collins, CO; aristoff@rams.colostate.edu}
\author{Dmitrii E. Makarov}
\affiliation{Department of Chemistry and Oden Institute for Computational Engineering and Sciences, University of Texas, Austin, TX; makarov@cm.utexas.edu}
\author{Gideon Simpson}
\affiliation{Department of Mathematics, Drexel University, Philadelphia, PA; grs53@drexel.edu}
\author{Daniel M. Zuckerman}
\affiliation{Department of Biomedical Engineering, Oregon Health and Science University, Portland OR; zuckermd@ohsu.edu}

\date{\today}

\begin{abstract}
Probability currents are fundamental in characterizing the kinetics of
non-equilibrium processes.  Notably, the steady-state current $\jss$ for a
source-sink system can provide the exact mean-first-passage
time (MFPT) for the transition from source to sink.  Because transient non-equilibrium behavior is quantified
in some modern path sampling approaches, such as the “weighted ensemble”
strategy, there is strong motivation to determine bounds on $\jss$ -- and hence on the MFPT -- as the system
evolves in time.  Here we show that $\jss$ is bounded from above and below by
the maximum and minimum, respectively, of the current as a function of the
\emph{spatial} coordinate at any time $t$ for one-dimensional systems undergoing
over-damped Langevin (i.e., Smoluchowski) dynamics and for higher-dimensional
Smoluchowski systems satisfying certain assumptions when projected onto a single
dimension.  These bounds become tighter with time, making them of potential
practical utility in a scheme for estimating $\jss$ and the long-timescale kinetics of complex systems.  Conceptually, the bounds
result from the fact that extrema of the transient currents relax
toward the steady-state current.
\end{abstract}


\maketitle


%
%

\section*{Introduction}

Non-equilibrium statistical mechanics is of fundamental importance in many
fields, and particularly in understanding molecular and cell-scale biology
\cite{hopfield1974kinetic,Hill2004,beard2008chemical,lee2019high}. Furthermore, fundamental theoretical ideas from the field
(e.g., Refs.\
\onlinecite{jarzynski1997nonequilibrium,crooks1999entropy,seifert2012stochastic})
have often been translated into very general computational strategies (e.g., Refs.\
\onlinecite{zhang2010weighted,Dickson2010a,ytreberg2004single,bello2015exact,nilmeier2011nonequilibrium}).

Continuous-time Markov processes occurring in continuous configurational spaces form a central pillar of non-equilibrium studies \cite{chapman1928brownian,kolmogoroff1931analytical}, including chemical and biological processes \cite{van1992stochastic}.
The behavior of such systems is described by a Fokker-Planck equation or, when momentum coordinates are integrated out, by the Smoluchowski equation \cite{Risken1996,gardiner2009stochastic}.
In the latter case, the probability density $p(x,t)$ and the probability current $J(x,t)$ are the key observables, and their behavior continues to attract theoretical attention \cite{berezhkovskii2014multidimensional,ghysels2017position,grafke2018numerical,polotto2018supersymmetric,cossio2018transition,del2018grand,berezhkovskii2018mapping}.
Continuous-time Markov processes in discrete spaces obey a master equation \cite{van1992stochastic}; such "Markov state models" play a prominent role in modern biomolecular computations \cite{singhal2004using,noe2007hierarchical,voelz2010molecular,chodera2014markov} as well as in the interpretation of experimental kinetic data \cite{makarov2015single,lee2019high}.

An application of great importance is the estimation of the mean first-passage time (MFPT)
for transitions between "macrostates" $A$ and $B$, two non-overlapping regions of configuration space.
The MFPT$(A \to B)$ is the average time for trajectories initiated in A according to some specified distribution, to reach the boundary of B.\cite{gardiner2009stochastic}
If a large number of systems are orchestrated so that trajectories reaching B (the absorbing ``sink'') are re-initialized in A (the emitting ``source''), this ensemble of systems will eventually reach a steady state.
Such a steady-state ensemble can be realized via the "weighted ensemble" (WE) path
sampling strategy
\cite{huber1996weighted,zhang2010weighted,suarez2014simultaneous,adhikari2018computational}, for example.
In steady state, the total probability arriving to state B per unit time (the ``flux'' -- i.e., the integral of the current $J$ over the boundary of B) will be constant in time.
The "Hill relation" then provides an exact relation between the
steady-state flux
and the mean first-passage time (MFPT)
\cite{Hill2004,bhatt2010steady,suarez2014simultaneous}:
\begin{equation}
\frac{1}{\mfpt(A \to B)} = \flux(A \to B \, | \, \mbox{steady state}) \; ,
    \label{hill}
\end{equation}
where the dependence on the initializing distribution in A is implicit and does not affect the relation.
When macrostates A and B are kinetically well-separated the reciprocal of the $\mfpt$ is the effective rate constant for the transition\cite{hanggi1990reaction}.
The Hill relation \eqref{hill} is very general \cite{Hill2004,warmflash2007umbrella,dickson2009nonequilibrium,aristoff2018optimizing,zuckerman2015statistical} and is not
restricted to one dimensional systems or to particular distributions of feedback
\emph{within} the source state A.

In one dimension, the steady flux is equivalent to the steady-state current $\jss$, which will be used in much of the exposition below.

While the relaxation time to converge to steady-state can be very short compared to the $\mfpt$, it is unknown a priori, and may be computationally expensive to sample in complex systems of interest \cite{adhikari2018computational}. 
This limitation applies to the weighted ensemble strategy, as traditionally implemented, because WE is unbiased in recapitulating the time evolution of a system \cite{zhang2010weighted}.
Thus there is significant motivation to obtain information regarding the converged steady-state current (which depends upon the boundary conditions, and \emph{not} the initial condition) of complex systems, from the observed transient current (which does depend upon the initial condition).
Here, we take first steps toward this goal and show that the maximum (minimum) transient current, regardless of initial condition, serves as an upper (lower) bound on the non-equilibrium steady-state current, in a class of one-dimensional continuous-time Markovian stochastic systems, with prescribed boundary conditions.

In cases of practical interest, the full system is likely to be high-dimensional, and we do not expect that the \emph{local} transient currents at particular configurations will provide a bound on the steady-state flux. 
However, one-dimensional (1D) projections of the current, along a well-chosen collective coordinate of interest, may still exhibit monotonic decay as discussed below.

The paper is organized as follows.
After proving and illustrating bounds in discrete and continuous 1D systems, we discuss more complex systems and provide a numerical example of folding of the NTL9 protein using WE\cite{adhikari2018computational}.  
We speculate that effective upper and lower bounds for the current exist in high-dimensional systems, and hope that the 1D derivations presented here will motivate future work in this regard.

\section*{Discrete-state formulation: Bounds and Intuition}

\begin{figure}
    \centering
    \includegraphics[width=200pt]{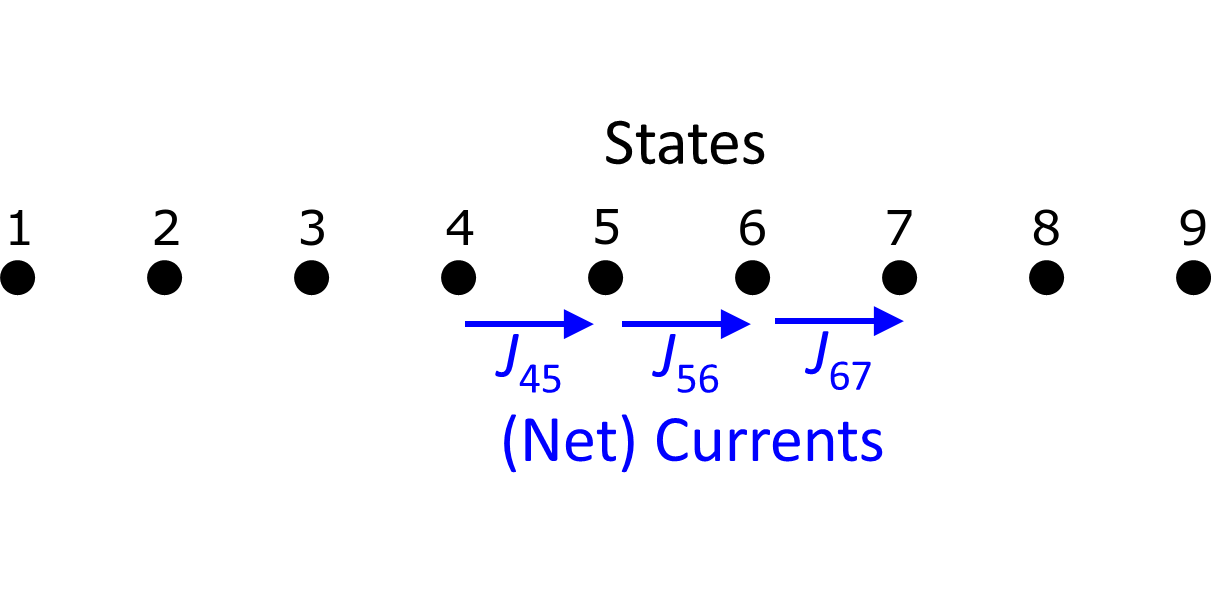}
    \caption{One-dimensional discrete state system.  States (black numbers) and currents (blue) are shown}
    \label{fig:discrete}
\end{figure}

The essence of the physics yielding bounds on the steady current can be appreciated from a one
dimensional continuous-time discrete-state Markov process as shown in Fig.\
\ref{fig:discrete}. The dynamics will be governed by the usual master equation
\begin{equation}
\frac{dP_i}{dt} = - \sum_{j \neq i} P_i \, \kij + \sum_{j \neq i} P_j \, \kji
\label{master}
\end{equation}
where probabilities $P_i = P_i(t)$ vary in time while rate constants $\kij \equiv k_{i,j}$ for $i \to j$ transitions are time-independent.
(We use a subscript convention throughout where the forward direction is left-to-right, and commas are omitted when possible.)
We will assume that only nearest-neighbor transitions are allowed -- i.e., that
\begin{equation}
\kij = 0 \hspace{2mm} \mbox{for} \hspace{2mm} |j-i|>1 \; .
\label{locality}
\end{equation}
Indeed, discrete random walks of this type provide a finite-difference approximation to diffusion in continuous space \cite{metzler2000random}. 
The net current in the positive direction between any neighboring pair of states is given by the difference in the two directional probability flows,
\begin{equation}
\jip = P_i \, \kip - P_{i+1} \, \kim
\label{jdef}
\end{equation}
Because the probabilities $P_i$ vary in time, so too do the currents: $\jip = \jip(t)$.
Using \eqref{jdef}, the master equation \eqref{master} can be re-written as
\begin{equation}
\frac{dP_i}{dt} = \jim - \jip
\label{masterj}
\end{equation}
which is merely a statement of the familiar continuity relation: the probability of occupying a state increases by the difference between the incoming and outgoing currents.

To establish a bound, assume without loss of generality that a local maximum of the current occurs between states 5 and 6.
That is,
\begin{equation}
    \jjk > \jij \hspace{1cm} \mbox{and} \hspace{1cm} \;
    \jjk > \jkl \; .
    \label{jassume}
\end{equation}
Differentiating $\jjk$ with respect to time and employing Eqs.\ \eqref{jdef} and \eqref{masterj} yields
\begin{align}
    \frac{d \jjk}{dt} &= k_{56} \frac{d P_5}{dt} - k_{65} \frac{d P_6}{dt} \nonumber \\
    &= k_{56} ( \jij - \jjk ) - k_{65} ( \jjk - \jkl ) < 0 \; ,
    \label{djdtmax_discrete}
\end{align}
where both terms are negative because rate constants are positive and the signs of the current differences are determined by the assumptions \eqref{jassume}.
The local maximum current must decrease in time.

If instead $J$ exhibited a local \emph{minimum} at the $5\to6$ transition, reversing the directions of the inequalities \eqref{jassume}, then the corresponding time derivative would be positive, implying a local minimum must increase.

We have therefore shown \emph{regardless of boundary conditions} that local maxima must decrease with time and local minima must increase with time in the discrete-state case with nearest-neighbor transitions.
Under stationary boundary conditions, the current will decay to its steady value $\jss$ and thus the global extrema at any time are bounds on the steady current.
Physically, the changes in the probability produced by local differences in current -- Eq.\ \eqref{masterj} -- necessarily cause relaxation of the current toward its steady value.
We note that in a one-dimensional steady state, whether equilibrium or non-equilibrium, the current $\jss$ is a constant independent of position.

\subsection*{Boundary behavior in a discrete-state source-sink system}
The preceding conclusions were  obtained for \emph{local extrema} without any
assumptions about boundary conditions. We now want to examine boundary
conditions of particular interest, namely a feedback system with one absorbing
boundary (``sink'') state and one emitting boundary (``source'') state, where flux
into the sink is re-initialized.  In such a source-sink system, we will see that similar
conclusions are reached regarding the relaxation of current extrema at the
boundaries.

For concreteness, suppose in our linear array (Fig.\ \ref{fig:discrete}) that
state 0 is the source and state 9 is the sink: the probability current into state
9 is fed back to state 0. The source at state 0 is also presumed to
be the left boundary of the system, which is implicitly reflecting.

Consider first the case where the maximum of the current occurs at the source at
some time $t$ -- i.e., $\jsource$ is the maximum. To analyze this case, note
that by assumption, the source state 0 in fact receives probability that arrives
to the sink state 9. That is, Eq.\ \eqref{masterj} applies in the form
\begin{equation}
    \frac{ d \psource }{ dt } = \jsink - \jsource \; .
\end{equation}
This, in turn, implies that the analog of Eq.\ \eqref{djdtmax_discrete} applies
directly, and we deduce that if $\jsource$ is the maximum among $J$ values, then
it must decrease in time.

Because analogous arguments apply to all the other boundary cases (maximum at
sink, minimum at either boundary), we conclude that any boundary extremum
current must decay with time toward $\jss$ in a source-sink discrete-state
system.

For completeness, we note that in principle the feedback of the flux reaching the
sink state could occur at a set of source states, in contrast to the single
source state assumed above. However, because of the locality property
\eqref{locality} which has been assumed, if we consider any current maximum not
part of the set of source states, the same arguments will apply.

\section*{Current bounds for continuous systems in the Smoluchowski framework}

Our primary interest is continuous systems, and so we turn now to a formulation
of the problem via the Smoluchowski equation, which describes over-damped
Langevin dynamics \cite{Risken1996,gardiner2009stochastic}. Conceptually,
however, it is valuable to note that the preceding discrete-state derivation of
current bounds depended on the \emph{locality} embodied in \eqref{locality} and the \emph{Markovianity} of
dynamics, two properties that are preserved in the Smoluchowski picture.

Our derivation proceeds in a straightforward way from the one-dimensional
Smoluchowski equation. Defining $p(x,t)$ and $J(x,t)$ as the probability density
and current at time $t$, we write the Smoluchowski equation as the continuity relation
\begin{equation}
    \frac{ \dee p }{\dee t} = - \frac{ \dee J }{\dee x}
    \label{cont}
\end{equation}
with current given by
\begin{equation}
    J(x,t) = \frac{ D(x) }{ k_B T } f(x) \, p(x,t) - D(x) \, \frac{ \dee p }{ \dee x } \; ,
    \label{jxt}
\end{equation}
where $D > 0$ is the (possibly) position-dependent diffusion ``constant'', $k_B
T$ is the thermal energy at absolute temperature $T$, and $f = -dU/dx$ is the
force resulting from potential energy $U(x)$. 

We now differentiate the current with respect to time and examine its behavior
at extrema -- local minima or maxima. We find
\begin{align}
    \frac{ \dee J }{\dee t}
    &= \frac{ D(x) }{ k_B T } f(x) \frac{ \dee p }{\dee t}
    -  \frac{ \dee }{\dee t} \left [ D(x) \, \frac{ \dee p }{ \dee x } \right] \\
    &= \frac{ D(x) }{ k_B T } f(x) \frac{ \dee p }{\dee t}
    -  D(x) \frac{ \dee^2 p }{ \dee x \dee t } \\
    &= -\frac{ D(x) }{ k_B T } f(x) \frac{ \dee J }{\dee x}
    + D(x) \, \frac{ \dee^2 J }{ \dee x^2 }  \label{djdt}
    \hspace{0.75cm} \mbox{[General]}\\
    &= D(x) \, \frac{ \dee^2 J }{ \dee x^2 }
    \hspace{2.75cm} \mbox{[Extrema only]}   \label{djdtmax}
\end{align}
where the third line is derived by equating  $ \dee^2 p / \dee t \dee x = \dee^2
p / \dee x \dee t$ and then substituting for $\dee p / \dee t$ in all three
terms using the continuity relation \eqref{cont}. The last line is obtained
because $\dee J / \dee x = 0$ at a local extremum (in $x$).

Eq. \eqref{djdtmax} is the sought-for result: it implies decay with
time of local extrema in the current $J$. If $x$ is a local maximum, then
$\dee^2 J / \dee x^2 < 0$ and conversely for a minimum; recall that $D(x)$ is
strictly positive. (Strictly speaking, for a local maximum one has $\dee^2 J /
\dee x^2 \leq 0$ rather than a strict inequality, but the case of vanishing second derivative is pathological for most physical systems.) 
With open boundaries, the global maximum is also a local maximum and must decrease in time. Likewise, the global minimum must increase. The global extrema therefore provide upper and lower bounds at any given $t$ that tighten with time. 
See below for discussion of boundaries and source/sink systems.

It is interesting to note that Eq.\ \eqref{djdt} resembles a Smoluchowski
equation, but for the current $J$ instead of $p$. Except in the case of simple
diffusion [$f(x) = 0$ and $D(x) = D = \mbox{const.}$], this is a ``resemblance''
only, in that the right-hand side cannot be written as the divergence of an
effective current and hence the integral of the current is not a conserved
quantity. However, the similarity may suggest why the current has a ``self
healing'' quality like the probability itself -- i.e., the tendency to relax
toward the steady state distribution.

\subsection*{Maximum Principle in a spatially bounded region}

The preceding results could be obtained with elementary calculus, but characterizing current extrema in systems with more challenging boundary behavior requires the use of mathematical approaches not well known in the field of chemical physics.
Mathematically, it is known that Eq.\
\eqref{djdt} is a ``uniformly parabolic'' equation -- defined below -- and therefore obeys a ``maximum principle'' (MP).\cite{protter2012maximum}
The MP, in turn, implies the monotonic decay of extrema noted above, away from boundaries.
In addition to vanishing first derivatives at extrema, the MP only requires the
\emph{non-strict} inequality $\dee^2 J / \dee x^2 \leq 0$, or the corresponding
inequality for a minimum.
For reference, we note that a uniformly parabolic partial differential equation for a function $u(x,t)$ takes the form
\begin{equation}
    \label{parabolic}
    \frac{ \partial u }{ \partial t} = a(x,t) \, \frac{ \partial^2 u }{ \partial x^2 } + b(x,t) \, \frac{ \partial u }{ \partial x} \,
\end{equation}
where $a(x,t) \geq a_0 > 0$ for all $x$ and $t$ in the domain of interest, with $a_0$ a constant.
Note that $b(x,t)$ is not restricted to be positive or negative.

The maximum principle dictates that if one considers the space-time plane
defined by $0 \leq x \leq L$ and $t_1 \leq t \leq t_2$, then any local extremum
must occur on the spatial boundaries ($x=0$ or $x=L$) or at the initial time
$t_1$. Most importantly, the extremum \emph{cannot} occur at $t=\tmax$ away from
the boundaries. Because $t_1$ and $t_2$ are arbitrary, then one can inductively
consider decreasingly small $t_1$ values arbitrarily close to $t_2$ to infer
monotonic decay of extrema which occur away from the boundaries. We note that
non-rectangular space-time domains are covered by MPs to some extent
\cite{protter2012maximum}.

It is interesting that the Smoluchowski equation itself for $p(x,t)$ does \emph{not} generally take the form \eqref{parabolic} and hence may not obey a maximum principle.
The value of the maximum of $p$ could grow over time.
One example is the relaxation of an initially uniform distribution in a harmonic potential, which would develop an increasing peak at the energy minimum as equilibrium was approached.
The density satisfies a maximum principle in \emph{simple} (force-free) diffusive behavior \cite{protter2012maximum} -- which does conform to \eqref{parabolic} -- in which the density must spread with time.
The current, like the density in simple diffusion, tends toward a constant value in steady state -- even when there is a non-zero force.
\subsection*{Maximum principle for a continuous source-sink system}\label{section:max-source-sink}

The case of primary interest is a source-sink feedback system because, as noted above, the steady current quantitatively characterizes the system's kinetics.
This 1D current is exactly the inverse MFPT, from the Hill relation \eqref{hill}.  
We have not yet explicitly considered the addition of source and sink terms in the Smoluchowski equation Eq. \eqref{cont}.  With explicit inclusion of source-sink feedback 
in one dimension, we find that a paradigmatic system which obeys a maximum principle over the entire domain (including boundaries) is a finite interval with one end acting as a perfect sink and the other end being the source where flux from the sink is reintroduced. The source boundary is taken to be reflecting, so this is not a fully periodic system.

When the global maximum or minimum occurs at a boundary of a one-dimensional
source-sink system -- either at the sink or the other boundary --
additional consideration (beyond what is discussed above) is necessary because the condition $\dee J/\dee x =0$
generally will \emph{not} hold at the boundaries. However, as motivation, we
point out that the same continuity arguments employed above in the discrete case
apply in the continuous case as well, at least for the case of feedback to a
single source state at a boundary.
Intuitively, then, monotonic decay of extrema is again expected.


Mathematically, we start by considering a system bounded by an interval $0 \leq x \leq x_1$ with sink at $x=0$ and source location $\xsrc \in (0,x_1)$.
The source is initially located in the interior of the interval for mathematical simplicity but later will be moved (infinitely close) to the boundary.
The probability current reaching the sink is re-initialized at $x=\xsrc$, while the $x_1$ boundary is reflecting in this formulation. 
The governing equation therefore includes a source term:
\begin{equation}
    \frac{ \dee p(x,t) }{\dee t} = - \frac{ \dee J(x,t) }{\dee x} - J(x=0,t) \, \delta(x-\xsrc)
    \label{cont_src}
\end{equation}
with current given again by \eqref{jxt}, with sink boundary condition
$p(x=0,t) = 0$, and with reflecting boundary condition $J(x=x_1,t)=0$ to a model a finite domain with no probability loss. 
The negative sign preceding the source term $J(x\!=\!0,t) \, \delta(x-\xsrc)$ is necessary because current arriving to the sink (at the \emph{left} side of the interval) is negative by convention.
Note that \eqref{cont_src} is a special case where feedback occurs at a point;
more generally, instead of a delta function, an arbitrary distribution on the
domain could be used in the second term on the right-hand side: see Appendix \ref{app:src_genl}.

In Appendix \ref{app:src_genl}, we show that \eqref{cont_src} obeys a maximum principle regardless of the location of the source $\xsrc$ or the initial condition.  However, on its own, this maximum principle does not establish the sought-for monotonic decay of extrema because maxima and minima could still occur on the spatial boundaries, or at the source, with increasing time. Note that the maximum principle applies only to global extrema inside the domain.

We therefore turn to an alternative formulation that includes a boundary source \emph{implicitly} via boundary conditions without the source term of \eqref{cont_src}, and a more powerful maximum principle is also seen to hold. 
As shown in Appendix \ref{app:periodic}, by taking the limit of $\xsrc\to x_1$ (or, equivalently, $x_1\to \xsrc$), we obtain the standard Smoluchowski description of Eqs.\
\eqref{cont} and \eqref{jxt} with sink boundary condition $p(x=0,t) = 0$, 
\emph{along with an additional boundary condition -- the ``periodicity'' of current,} namely, $J(x=\xsrc, t) = J(x=0,t)$. 
The source term of Eq.\ \eqref{cont_src} is no longer present, but
identical behavior for $p$ and $J$ is obtained, along with a maximum principle, as shown in Appendix \ref{app:periodic}.


In this special case when the single source point occurs at a boundary, the
periodicity of the current 
does not allow a local extremum on the boundary and leads to a 
maximum principle (MP) implying monotonic decay of extrema; see Appendix \ref{app:periodic}. 
The MP for a periodic function indicates that the maximum in a space-time
domain between arbitrary $t_1 < t_2$ must occur at the earlier time. This
implies, inductively, the monotonic decay of local extrema in $J$ -- i.e.,
decrease with $t$ of maxima and increase of minima.

Although monotonic decay of extrema may seem obvious from the discrete case, the maximum principle for the continuous case covers instances that may seem surprising.
In looking for a counter-example, one could construct a system with a very low diffusion constant but large and spatially varying forces.
For example, one could initialize a spatially bounded probability distribution on the side of an inverted parabolic potential: intuitively, one might expect the maximum current to increase as the probability packet mean velocity increases down the steepening potential. 
However, so long as the diffusion rate is finite, the spreading of the probability distribution (lowering the peak density) will counteract the increase in velocity. A numerical example is shown in Figure \ref{fig:parabolaPotential}. 

\subsection*{Numerical evidence: One dimension}

\begin{figure}
    \centering
    \includegraphics[width=0.9\linewidth]{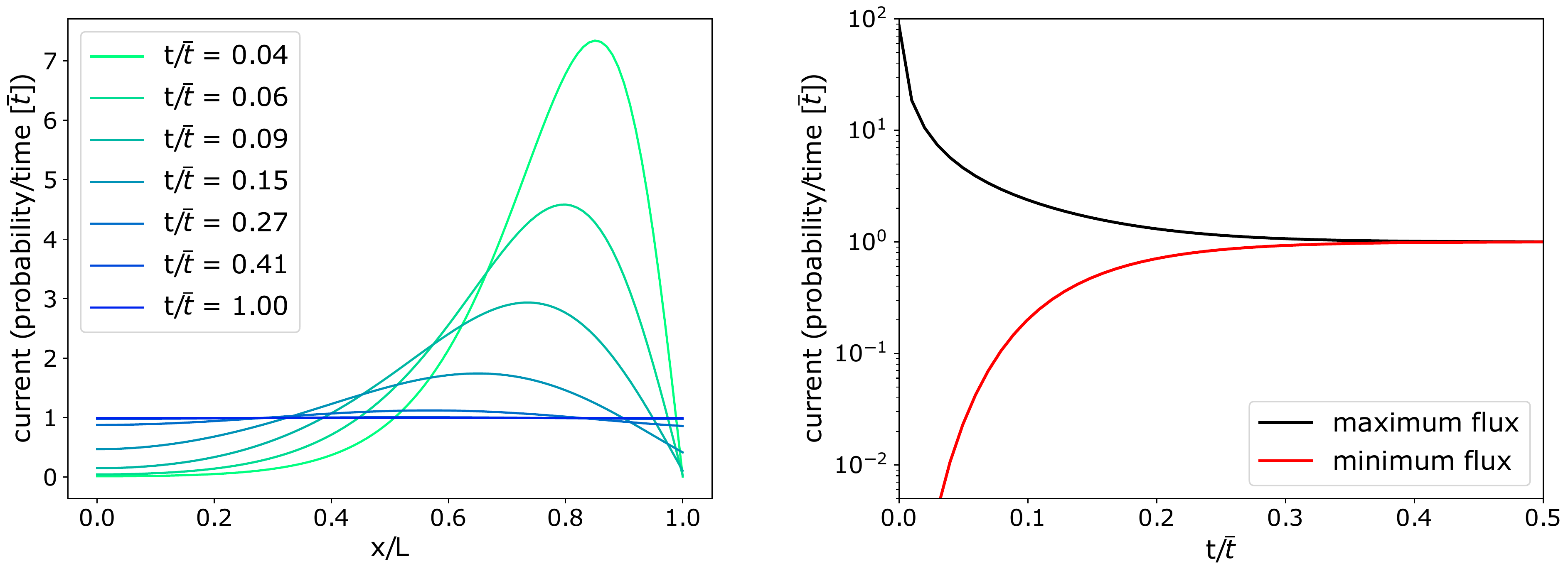}
    \caption{Numerical data for simple diffusion in a source-sink system.  Left: The system is initialized with all probability at the source $x/L=1$, and the current towards the sink is seen to relax to steady behavior as $t$ becomes a substantial fraction of the first-passage time $\bar{t}$.  Right: The maximum and minimum currents converge monotonically toward the steady-state value with increasing time.}
    \label{fig:simple}
\end{figure}

We have employed numerical analysis of one-dimensional systems to illustrate the
behavior of the time-evolving current. 
In these examples, we define positive current as directed towards the sink. 

We first examined a simple diffusion
process with source at $x=1$ and sink boundary condition at $x=0$ using units
where the diffusion coefficient $D=1$ and the mean first-passage time is
$\bar{t}=L^2/2D$, where $L=1$ is the domain length. In all examples probability is initialized as a delta-function distribution at the source, and propagated via numerical solution of the Smoluchowski equation using the FiPy package \cite{guyer2009fipy}.  In all examples we have applied periodic boundary conditions for the
current (with a reflecting boundary at the source), appropriate to describe the evolution of Eq.\
\eqref{cont_src} for a single-source point at the system boundary (see
Appendix \ref{app:periodic} for a complete discussion of boundary conditions).    
Fig. \ref{fig:simple} shows clearly that the spatial maximum and
minimum value bracket the true steady-state current. In this system the minimum current value and the ``target'' current (at the sink) are identical.

\begin{figure}
    \centering
    \includegraphics[width=0.9\linewidth]{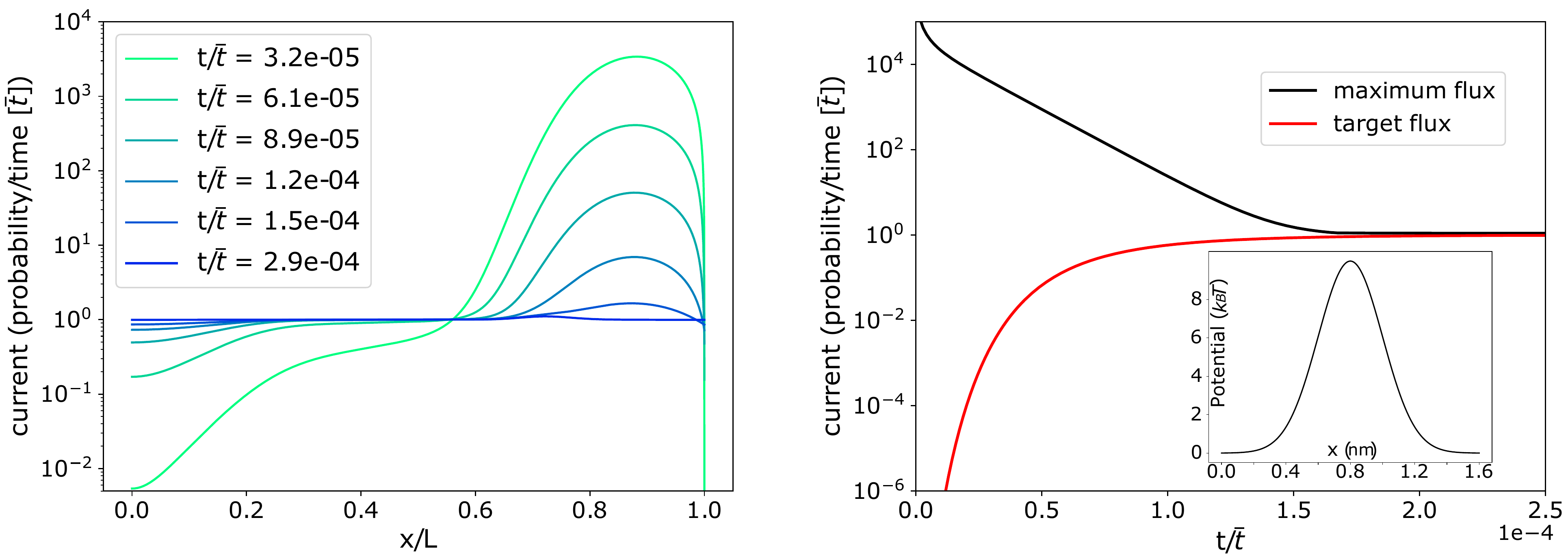}
    \caption{Numerical data for diffusion over a central barrier in a source-sink system.  Left: The system is initialized with all probability at the source, $x/L=1$, and the current towards the sink at $x=0$  is seen to relax to steady behavior at a fraction of the $\mfpt$. 
    Right: The currents at the maximum and sink (identical to the minimum in this system) converge monotonically toward the steady-state value with increasing time. Inset: Potential energy in the domain.}
    \label{fig:potential}
\end{figure}

We also examined a numerical system with a potential barrier separating the
source ($x=1.6$ nm) and sink ($x=0$) states.  See Fig.\ \ref{fig:potential}.
Parameters for this example were roughly intended to model the diffusion of a
1nm sphere in water at 298K: $D=2.6\times10^{-10}$ m$^2$/s, and a Gaussian
barrier of height 10 $k_B T$ and width 2 nm. Probability was again initialized at the source. Qualitatively the results are
similar to the simple diffusion case, with the spatial maximum and minimum current
bracketing the true steady-state current.
In this system the minimum current value and the current at the sink are identical.  
Here, the relaxation time to steady state is roughly four orders of magnitude faster than the $\mfpt$ of $\sim0.5$ ms, as shown in Fig. \ref{fig:potential}. 

\begin{figure}
    \centering
    \includegraphics[width=0.9\linewidth]{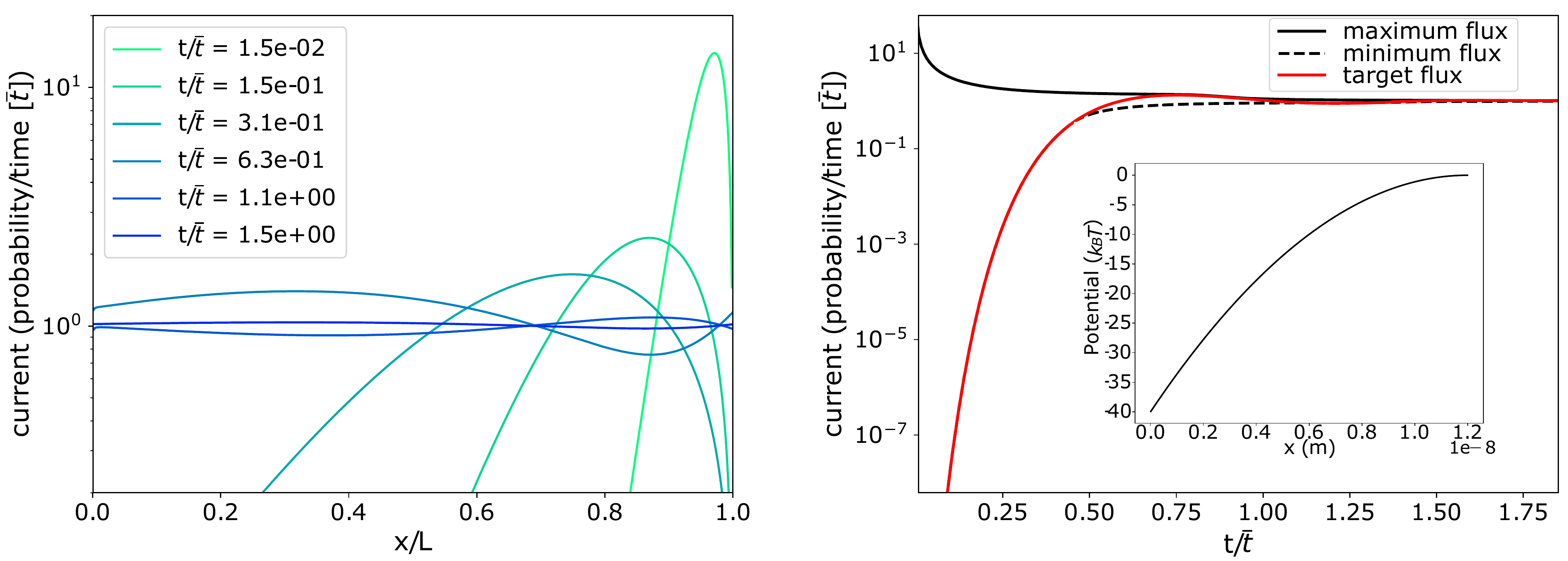}
    \caption{Numerical data for diffusion down an inverted parabola potential in a source-sink system ($\mfpt = \overline{t}=6.2$ns). Left: The system is initialized with a delta-distribution at the source, and the current towards the target is plotted from the source at $x=12.0$nm to the sink (target) at $x=0$, and is seen to relax monotonically to steady-state.  Right: The maximum (solid black) and minimum (dashed black) currents converge monotonically toward the steady-state value with increasing time, while the current at sink (red) relaxes non-monotonically. Inset: Potential energy in the domain.}
    \label{fig:parabolaPotential}
\end{figure}

Finally, we examined a one-dimensional in which the monotonic decay of the current may not be intuitive. 
We initialize a delta-function distribution at the top of an inverted parabolic potential $U(x)=-\frac{1}{2}k (x-x_0)^2$ and force constant $k=\frac{5 k_B T}{(3 \mbox{nm})^2}$, with the source at the peak ($x_0=12.0$ nm) and sink at $x=0$.
Dynamics parameters for this example are identical to the previous case, $T=$ 298K and $D=2.6\times10^{-10}$ m$^2$/s. 
Even though the mean velocity of the ``particle'' initialized at the top of the inverted parabola increases rapidly, this acceleration is counteracted by the spreading of the initial distribution.
In accordance with the maximum principle, Fig.\ \ref{fig:parabolaPotential} shows that the current maximum (minimum) monotonically decreases (increases) until steady-state is reached ($\mfpt=6.2$ns).  
Interestingly, in this system the minimum current value and the current at the sink differ. 
Although the maximum principle implies the minimum current will increase monotonically over time, the MP does not intrinsically characterize the target current (at the sink), which may not be a minimum.

\section*{Discussion of more complex systems}

Should we expect that analogous bounds exist in cases of practical interest,
when the current from a high-dimensional system is integrated over iso-surfaces of a
one-dimensional coordinate $q$? 
This is a situation often encountered in molecular simulation, where conformational transitions of interest require correlated motion between many hundreds or thousands of atoms, and are observed along a handful of collective coordinates.
In fact, there is no maximum principle for the locally defined current magnitude in higher dimensional spaces, but even when the local high-dimensional current magnitude does not monotonically decay, the flux over iso-surfaces of a reaction coordinate may exhibit monotonic decay: see below and Appendix \ref{app:2Dcounterexample}. 
We have not derived general results for this case, but there are interesting hints in the literature that a more general formulation may be possible. 

Most notably, Berezhkovskii and Szabo showed that the probability density
$p(\phi,t)$ of the ``committor" coordinate $\phi$ evolves according to a
standard Smoluchowski equation under the assumption that ``orthogonal
coordinates'' (along each iso-committor surface) are equilibrium-distributed
according to a Boltzmann factor \cite{berezhkovskii2013diffusion}; see also Ref.\ \onlinecite{lu2014exact}. Note that the
committor $0 \leq \phi \leq 1$ is defined in the full domain to be
the probability of starting at each point and reaching a chosen target state before visiting the given
initial state. Because our preceding derivation of current bounds for
one-dimensional systems relied entirely on the Smoluchowski equation, it follows
that the current projected onto the committor, $J(\phi)$ would be subject to the
same bounds -- so long as the additional assumption about equilibrated
orthogonal coordinates holds
\cite{vanden2003fast,hartmann2007model,legoll2010effective}.

\begin{figure}
    \centering
    \includegraphics[width=0.7\linewidth]{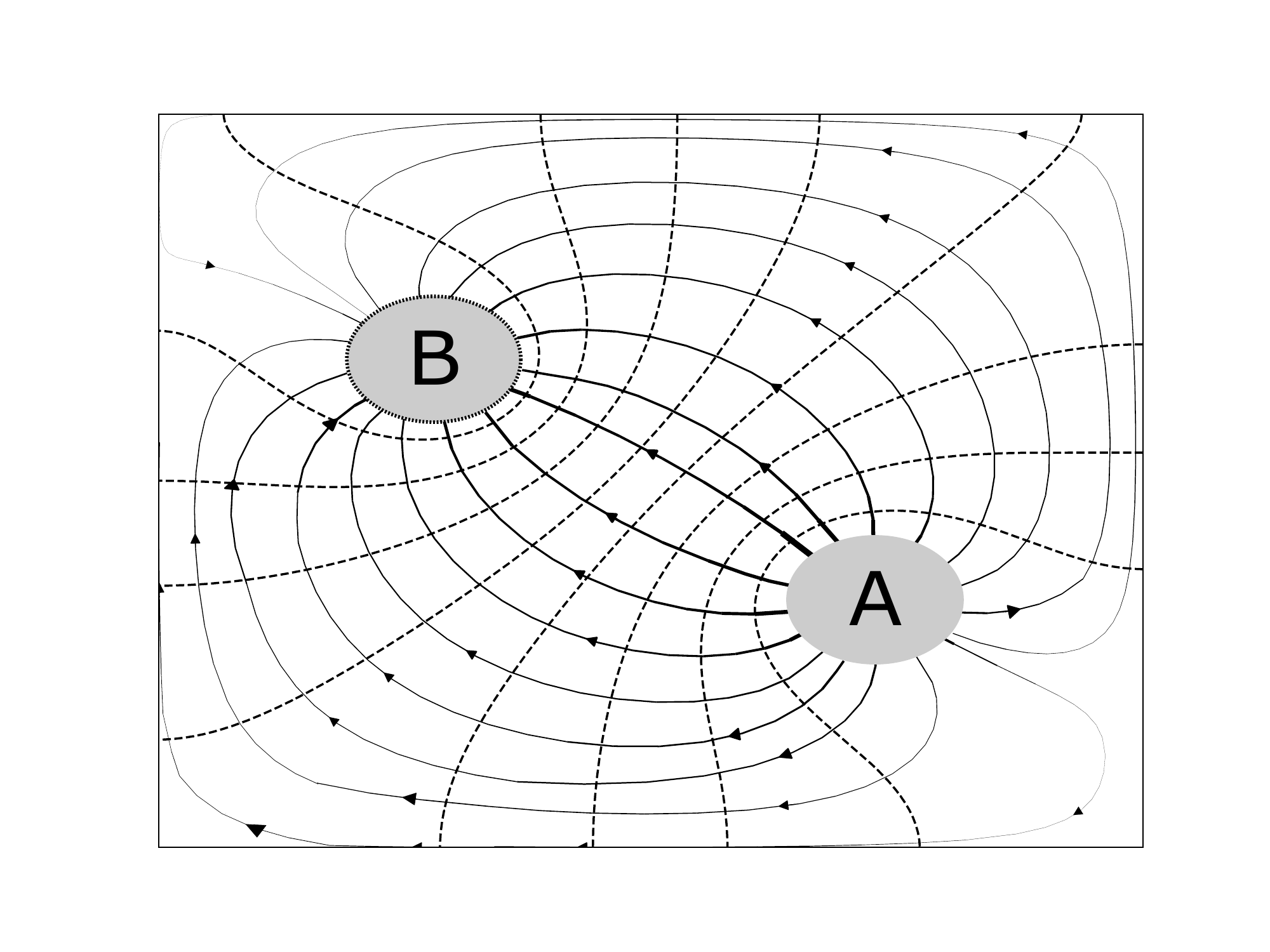}
    \caption{Schematic of flows and isocommitor surfaces.  Steady-state current flow lines (solid lines with arrows) and committor isosurfaces (dashed lines) are shown in a bounded domain with source (A) and sink (B) states.  As discussed in the text, no component of the steady-state current can flow along isocommittor surfaces, which also must exhibit the equilibrium distribution, when suitable boundary conditions are enforced.}
    \label{fig:fluxlines}
\end{figure}

It is intriguing to note that the orthogonal equilibration assumption is true in one type of A $\to$ B steady state. 
Consider a steady state constructed using "EqSurf" feedback \cite{bhatt2011beyond}, in which probability arriving to the target state B is fed back to the \emph{surface} of initial state A according to the distribution which would enter A (from B) in equilibrium; this procedure preserves the equilibrium distribution within A \cite{bhatt2011beyond}. 
For \emph{any} steady state, the current is a vector field characterized by flow lines, each of which is always tangent to the current.
Then, the probability density on any surface orthogonal to the flow lines must be in equilibrium: if this were not the case, a lack of detailed balance would lead to net flow of probability, violating the assumption of orthogonality to the current lines. A visual schematic of such a steady-state is shown in Figure \ref{fig:fluxlines}.
The same orthogonal surfaces must also be iso-committor surfaces in the EqSurf case, 
which can be shown by direct calculation.
Using the known relationship between the steady current, committor and potential energy for the EqSurf steady state \cite{berezhkovskii2013diffusion} one finds that the current is indeed parallel to the gradient of the committor:
\begin{equation}
\vec{\jss} = (1/Z) \, e^{-\beta U(\vec{x})} \, \nabla \phi(\vec{x}) \;, \end{equation}
where $\vec{x}$ is the full set of configurational coordinates, $\phi$ is the committor, and $Z$ is the system partition function.
This special case of ``orthogonal equilibration'' is quite interesting, but we remind readers that the transient (pre-steady-state) behavior orthogonal to current lines has not been characterized here.

\begin{figure}
    \centering
    \includegraphics[width=0.7\linewidth]{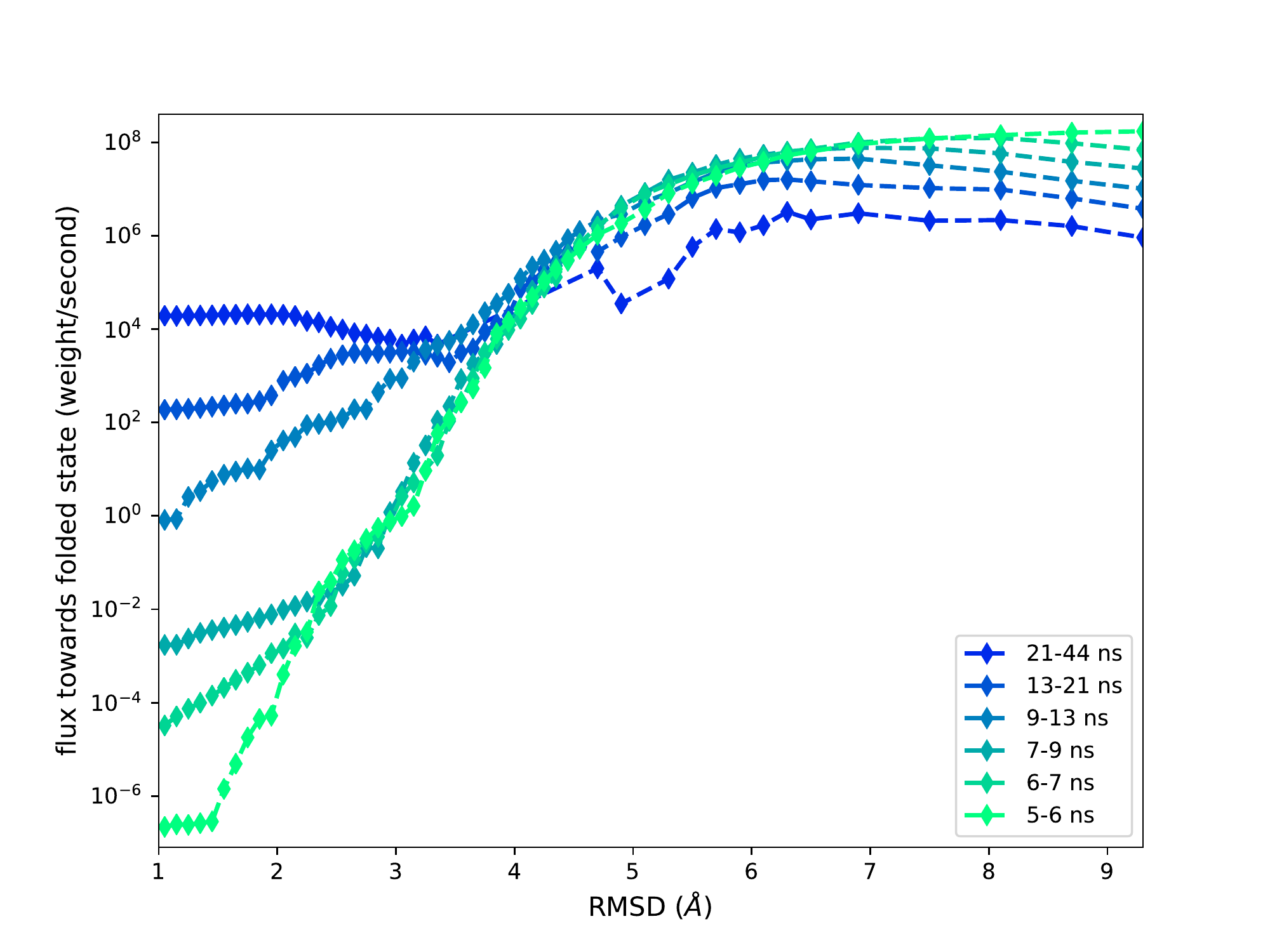}
    \caption{Numerical data for a complex system -- atomistic protein folding.
    The data show protein folding flux for atomistic, implicitly solvated NTL9 \cite{adhikari2018computational} as a function of a projected coordinate (RSMD) averaged over several time intervals during a simulation.
    The flux is the total probability crossing the indicated RMSD iso-surface per second.
    Data were obtained from weighted ensemble simulation, which orchestrates multiple trajectories to obtain unbiased information in the full space of coordinates over time -- i.e., Fokker-Planck-equation behavior is recapitulated \cite{zhang2010weighted}.
    Only positive (folding direction) current is shown, although some RMSD increments exhibit negative flux in some time intervals due to incomplete sampling/noise.}
    \label{fig:protein}
\end{figure}

We also provide numerical evidence for nearly monotonic relaxation behavior of the
current in a highly complex system, an atomistic model of the protein NTL9
undergoing folding. Fig.\ \ref{fig:protein} shows the \emph{flux} (total probability per second) crossing iso-surfaces of a collective variable, the RMSD, which here is the minimum root
mean-squared deviation of atom pair distances between a given configuration and
a fixed reference folded configuration -- minimized over all translations and
rotations of the two configurations. Since the collective variable iso-surfaces separate the folded and unfolded states, at steady state the flux will become constant across iso-surfaces of the collective variable. Data were harvested from a prior study
using the weighted ensemble (WE) approach, which was implemented with a source
at one unfolded configuration and a sink at the folded state, defined as RMSD
$\leq 1 \buildrel _{\circ} \over {\mathrm{A}}$ \cite{adhikari2018computational}.
Although the RMSD is a distance measure from an arbitrary configuration to the
folded state, it is \emph{not} claimed to be a proxy for the committor
coordinate described above.  Note that the WE method runs a set of unbiased
trajectories and performs occasional unbiased resampling in path space
\cite{zhang2010weighted}; thus WE provides the correct time-evolution of
currents and probability distributions, which are derived directly from the path
ensemble.  

Although the RMSD coordinate used in the NTL9 simulations is not likely to be an ideal reaction coordinate, we still observe monotonic relaxation of the flux profile. For this set of 30 weighted ensemble simulations of NTL9 protein folding, during the observed transient regime (where all trajectories were initialized in the unfolded state), the steady state is monotonically approached out to 45 ns molecular time (reflecting 225$\mu$s of aggregate simulation).
Further, although the current profile was still evolving in the NTL9 simulations and not fully steady, using the Hill relation \eqref{hill} for the $\mfpt$ from the flux into the folded state yielded a folding time of 0.2-2.0 ms, consistent with the experimental value\cite{adhikari2018computational}.    

Identifying good reaction coordinates to describe long-timescale conformational transitions remains a challenging problem in complex systems,\cite{best2005reaction,rohrdanz2013discovering,mcgibbon2017identification} and is beyond the scope of the present study.
For a perfect collective variable which captures the ``slow'' coordinates such that orthogonal degrees of freedom are equilibrated, the system can be effectively described by a 1D Smoluchowski equation and thus the global flux extrema will relax monotonically.  
Our hope is that this work, proving monotonic decay for the current in simple 1D systems, will inspire work to show how the projected current on imperfect reaction coordinates can provide bounds for the steady-state current.

In the realm of speculation, motivated in part by our numerical data, one can
ask whether a variational principle should hold. That is, if there are projected
coordinates with higher and lower current maxima, is the lower maximum always a
valid bound? This is a question for future investigation.

\section*{Implications for Numerical Computations}
There is significant motivation for pursuing steady-state behavior: in the non-equilibrium source-sink feedback systems studied here, the steady-state flux yields the mean first-passage time (MFPT) for transitions between two macrostates of interest via the Hill relation Eq. \eqref{hill}.  
The relaxation time to steady state can be many orders of magnitude faster than the $\mfpt$ in kinetically well-separated systems without significant intermediate metastable states: see Fig.\ \ref{fig:potential}.  Hence, large gains in estimating the MFPT could be obtained by sampling from the short-time non-equilibrium trajectory ensemble if the flux can be tightly bounded from above and below.  
Such transient information has been leveraged in Markov State Modeling approaches\cite{singhal2004using,noe2007hierarchical,chodera2014markov}, but a lack of separation of timescales between individual states can bias kinetic predictions\cite{suarez2016accurate}. When sampling from the non-equilibrium trajectory ensemble, the concerns are different. We propose that the observed transient flux can bound the steady-state flux along suitable reaction coordinates \cite{berezhkovskii2013diffusion} with clear separation of timescales between ``slow'' reaction progress and orthogonal degrees of freedom -- that is, in systems where transitions which are effectively one-dimensional.  These bounds will become tighter as sampling time increases; the $\mfpt$ is estimated exactly via \eqref{hill} when the sampling converges to steady state.


In terms of practical estimators, having upper and lower bounds based on the
``spatial'' variation of the flux would imply that \emph{any} spatial average
of the flux is a valid estimator for the steady flux that must
converge to the true value at large $t$. For a high-dimensional system, the
``spatial'' average would correspond to an average along the collective
coordinate exhibiting the bounds. Such an average could be linear or non-linear.

The potential value of such average-based estimators can be seen from the
 spatial current (flux) profiles plotted in Figs.\ \ref{fig:simple}-\ref{fig:protein}, where the
$\mfpt$ is estimated based on the flux into the sink
and is essentially the minimum flux during the transient period prior to steady state.
It seems clear that averaging among values
between the minimum and maximum would yield, at moderate $t$, values much closer
to the steady flux reached after long times.
Such estimators will be explored in the future.

\section*{Acknowledgements}

We are very appreciative of discussions with Sasha Berezhkovskii and David Zuckerman.
This work was supported by NSF Grants CHE-1566001 (to DEM), DMS-1522398 (DA) and DMS-1818726 (DA and GS), by NIH Grant GM115805 (DMZ), as well as by the Robert A. Welch Foundation Grant No. F-1514 (DEM).



\appendix

\section{Maximum Principle for a Source-Sink System with General Source Term}
\label{app:src_genl}


In this appendix, we derive a non-standard maximum principle for a general source-sink system. 
Recall that we consider a system on the  spatial interval $0 \le x \le x_1$ with a sink at 
$x=0$. We introduce a source $\gamma = \gamma(x)$ corresponding to a probability density function  with  derivative $\gamma'$. We suppose $\gamma$ is defined on an interval away from 
the source and sink; thus $\gamma(x) = 0$ for $x < a$ or $x > b$, and $a,b$  satisfy $0<a<b<x_1$.   It is not essential that there be a boundary at $x_1$; indeed, the arguments below apply just as well on $0 \le x < \infty$.  We retain the notation from the main manuscript for consistency. Let $p = p(x,t)$ be the probability solving
\begin{equation}
\frac{\dee p}{\dee t} = -\frac{\dee J}{\dee x} - \gamma(x) \cdot J(x = 0), \qquad 
p(x = 0) = 0, \quad J(x=x_1) = 0 \;.
\label{appsrc}
\end{equation}
$J$ is the current defined in \eqref{jxt}, and the overall negative sign of the source term is simply due to the overall negative direction of the current given the sink at $x=0$.

To obtain the maximum principle for the current, we first differentiate \eqref{jxt} with respect to time $t$ and use \eqref{appsrc} to obtain
\begin{align}
\begin{split}
\label{Jmax}
\frac{\dee J}{\dee t} &=
-\frac{ D(x)}{k_B T}f(x) \frac{\dee J}{\dee x}
    + D(x)\frac{\dee^2 J}{\dee x^2} + \left(-\frac{D(x)}{k_B T} f(x)\gamma(x) + \gamma'(x)D(x)\right)J(x = 0) 
\end{split}
\end{align}
which differs from~\eqref{djdt} due to the source terms. 
Let $\Omega_T$ be the set in the space-time domain 
corresponding to 
$(x,t)$ with $0<t\le T$ and either $0<x<a$ or $b<x<x_1$. 
We claim a (weak) maximum principle holds for $J$ in $\Omega_T$.

Let ${\bar \Omega}_T$ consist 
of $(x,t)$ with $0 \le t \le T$ and either $0 \le x \le a$ or $b \le x \le x_1$.
A maximum principle will imply that the maximum of $J$ over 
${\bar \Omega}_T$ must be attained outside of $\Omega_T$. This means the maximum 
of $J$ over $0 \le t \le T$ and $0 \le x \le x_1$ must 
be attained either at $t=0$, or at $t>0$ with $x=0$, $x = x_1$, or 
$a \le x \le b$. In other words the maximum current 
occurs either at time zero, or on the spatial boundary, 
 or in the source states.
We prove this by showing that a contradiction is reached otherwise, as 
follows.

\begin{itemize}
\item Let $(x_0,t_0)$ be a point where $J$ achieves its maximum in ${\bar \Omega}_T$.
    \item Suppose, for a moment, the maximum is nondegenerate,  $0< t_0 < T$, and $0<x_0<a$ or $b<x_0<x_1$.  By nondegenerate, we mean that $\frac{\dee^2 J}{\dee x^2}(x_0,t_0) < 0$.  Since it is a maximum,  
    $\frac{\dee J}{\dee x}(x_0,t_0) = 0$. Moreover $\gamma(x_0) = \gamma'(x_0) = 0$ as $\gamma(x) = 0$ for $x< a$ or $x> b$.  But this  contradicts~\eqref{Jmax}, since we are left with $0 =
    \frac{\dee J}{\dee t}(x_0,t_0) = D(x_0) \frac{\dee^2 J}{\dee x^2}(x_0,t_0)<0$.
    
    \item  Now suppose the maximum is nondegenerate with $t_0=T$, and $0 < x_0 < a$ or $b < x_0 < x_1$. Then $\frac{\dee J}{\dee t}(x_0,T) \ge 0$, $\frac{\dee J}{\dee x}(x_0,T) =\gamma(x_0) = \gamma'(x_0) =  0$, and $\frac{\dee^2 J}{\dee x^2}(x_0,T)<0$.  Again, we have a contradiction since  
    $0\leq \frac{\dee  J}{\dee t} (x_0,T) =  D(x_0)\frac{\dee^2 J}{\dee x^2}(x_0,T)<0$.

    \item In general, the maximum may be degenerate, and we make the 
    following perturbative argument. Let $\epsilon>0$ and define $$J^\epsilon(x,t) = J(x,t)-\epsilon t.$$ Then
\begin{equation}
\label{ueps}
\frac{\dee J^\epsilon}{\dee t} = -\epsilon -\frac{ D(x)}{k_B T}f(x)\frac{\dee J^\epsilon}{\dee x} 
    + D(x)\frac{\dee^2 J^\epsilon}{\dee x^2} + \left(-\frac{D(x)}{k_B T} f(x)\gamma(x) + \gamma'(x)D(x) \right)(J^{\epsilon}(x = 0) +\epsilon t).
\end{equation}
Consider ${\bar \Omega}_T\setminus \Omega_T$, 
defined as all space-time points in ${\bar \Omega}_T$ that are 
not in $\Omega_T$. Let $M = \max_{\bar \Omega_T \setminus \Omega_T} J$ be the maximum of $J$ over  ${\bar \Omega}_T\setminus \Omega_T$. We claim that the maximum of $J$ over 
${\bar \Omega}_T$ is less than $M$, that is,
$\max_{\bar \Omega_T}J \le M$. To prove this, 
we will show instead that the 
max of $J^\epsilon$ over ${\bar \Omega}_T$ is less than $M$, 
that is, $\max_{\bar \Omega_T}J^\epsilon \le M$. Once 
the latter is established, we have $\max_{\bar \Omega_T}J \le M+\epsilon T$ and the result follows by letting $\epsilon \to 0$.

First, notice that $\max_{\bar \Omega_T\setminus \Omega_T}J^\epsilon \le M$, since by definition of $M$ we have $J^\epsilon(x,t)  = J(x,t)-\epsilon t \le M-\epsilon t \le M$ for any $(x,t)$ in ${\bar \Omega}_T\setminus \Omega_T$. If $J^{\epsilon}(x,t)$ has a maximum 
at $(x_0,t_0)$ in $\Omega_T$, then $\frac{\dee J^\epsilon}{\dee t}(x_0,t_0) \ge 0$ while $\frac{\dee J^\epsilon}{\dee x}(x_0,t_0) = 0$, $\frac{\dee^2 J^\epsilon}{\dee x^2}(x_0,t_0) \le 0$, and $\gamma(x_0) = \gamma'(x_0) = 0$. But this contradicts~\eqref{ueps}, since
$$
0\le \frac{\dee J^\epsilon}{\dee t}(x_0,t_0) = - \epsilon  + D(x_0) 
\frac{\dee^2 J^\epsilon}{\dee x^2}(x_0,t_0) < 0.
$$
Therefore, the maximum of $J^\epsilon$ over $\bar \Omega_T$ must occur on 
${\bar \Omega}_T\setminus \Omega_T$ and $\max_{\bar \Omega_T} J^\epsilon = \max_{\bar \Omega_T\setminus \Omega_T} J^\epsilon \le M$, as desired.

\item The exact same arguments can be applied
using ``minimum'' in place of ``maximum,''
where in the last step above $\epsilon$
would be replaced with $-\epsilon$.
\end{itemize}

In  Appendix \ref{app:periodic}, due to the periodic  boundary condition and the source being at the 
boundary, we conclude that the maximum of $J$ is attained at the initial time. This 
leads the result cited in the main  text above, that the time dependent current gives bounds for
the steady state current $J_{ss}$.  Note that this is not true in the setting of this
Appendix, since  for a general source  $\gamma$, the maximum may be attained after 
the initial time in either the source states or on the spatial boundary.

\section{Maximum principle for a source-sink system with point source at the boundary}
\label{app:periodic}

Here, we show that a source-sink system with a point source 
at the boundary $x_1$ and a sink at $0$ 
satisfies a maximum principle, and that 
the extrema of the time dependent current give bounds 
for the steady state current $J_{ss}$. We 
begin with a special case of \eqref{appsrc} that corresponds to a source-sink 
system on $0 \le x \le x_1$ with a point source at 
$\xsrc$, sink at $0$, and reflecting boundary 
at $x = x_1$:
\begin{equation}\label{1}
\frac{\dee p}{\dee t} = -\frac{\dee J}{\dee x} - \delta(x-\xsrc) J(x=0), \quad p(x=0) = 0,\, \,J(x=x_1) = 0,
\end{equation}
where we assume $0 < x_{src} < x_1$. 

We first show 
that when $x_1 \to \xsrc$, the PDE in \eqref{1} is, 
in a sense, equivalent to the ordinary continuity relation (no source term) but different boundary conditions, namely,
\begin{equation}\label{2}
\frac{\dee \tilde{p}}{\dee t} = -\frac{\dee \tilde{J}}{\dee x}, \quad \tilde{p}(x=0) = 0,\,\,\tilde{J}(x=0) = \tilde{J}(x=\xsrc) \; .
\end{equation}
Above, $J$ is given by \eqref{jxt}, and $\tilde{J}$ is analogously defined, with $\tilde{p}$ taking the place of $p$. Equation  \eqref{1} is posed on $0 \le x \le x_1$ while \eqref{2} is posed on $0 \le x \le \xsrc$. Equation \eqref{2} thus corresponds to 
a source-sink system with sink at $0$ 
and a reflecting boundary and point source at $x_1$. 
Since \eqref{2} is identical to \eqref{djdt} but 
with periodic boundary condition on the current, we 
obtain a maximum principle for $J$ as well as monotonic 
convergence of the extrema of the time dependent current 
toward its steady state value $J_{ss}$, 
as discussed below.
An intuitive argument supporting current periodicity is given at the end of this appendix.

%


To address the limit $x_1 \to \xsrc$, we consider the ``weak solutions'' associated with \eqref{1} and \eqref{2}; weak solutions are a common framework for studying partial differential equations \cite{evans10}.   The weak solutions are
obtained by multiplying by a smooth test function and then
integrating by parts.  For \eqref{1}, if $\phi = \phi(x)$ is a smooth function on
$0 \le x \le x_1$ vanishing at $x=0$ with derivative $\phi'$,
\begin{align}
    \begin{split}
        \label{comp1}
        \int_0^{x_1} \frac{\dee p}{\dee t} \phi\,dx &=- \int_0^{x_1}  
        \frac{\dee J}{\dee x}\phi\,dx - \int_0^{x_1} \delta(x-\xsrc) J(x=0)\phi\,dx \\
        &= - \int_0^{x_1} \frac{\dee J}{\dee x} \phi\,dx - J(x=0)\phi(\xsrc)\\
        &= J(x=0)\phi(0)-J(x=x_1)\phi(x_1) -J(x=0)\phi(\xsrc)+ \int_0^{x_1} \phi' J\,dx \\
         &= -J(x=0)\phi(\xsrc)+ \int_0^{x_1} \phi' J\,dx .
\end{split}
\end{align}
In parallel, for \eqref{2}, we multiply by $\tilde{\phi}$, a smooth function on
$0 \le x \le  \xsrc$ vanishing at $x=0$ with derivative $\tilde{\phi}'$,
\begin{align}
    \begin{split}
        \label{comp2}
        \int_0^{\xsrc} \frac{\dee \tilde{p}}{\dee t} \tilde\phi\,dx&= -\int_0^{\xsrc} \frac{\dee \tilde{J}}{\dee x}\tilde\phi\,dx \\
        &= \tilde{J}(x = 0)\tilde\phi(0) -\tilde{J}(x = \xsrc)\tilde\phi(\xsrc)+ \int_0^{\xsrc} \tilde{\phi}'\tilde{J}\,dx\\
        &=   -\tilde{J}(x = 0)\tilde\phi(\xsrc)+ \int_0^{\xsrc} \tilde{\phi}'\tilde{J}\,dx.
\end{split}
\end{align}
 Thus the PDEs~\eqref{1} and~\eqref{2} have the same weak solutions
 when $x_1\to \xsrc$.


We can now obtain a maximum principle on the current. Dropping the $\tilde{\phantom{a}}$'s and differentiating~\eqref{2} yields \eqref{djdt} together with the boundary conditions
\begin{equation}
\label{e:bc1}
    J(x = 0) = J(x =\xsrc), \quad \frac{\dee J}{\dee x}(x=0) = 0.
\end{equation}
The last boundary condition comes from $p(x = 0) = 0$. Indeed $p(x = 0) = 0$  implies $\frac{\dee J}{\dee x}(x = 0) = -\frac{\dee p}{\dee t}(x = 0) = 0$.

With the periodic boundary condition on the current, a  version of the maximum principle applies, 
and shows that the max of $J$
over the time-space domain occurs at the initial time:
\begin{equation*}
    \max_{t_0\le t \le T, 0 \le x \le \xsrc}J(x,t) = \max_{0 \le x \le \xsrc}J(x,t_0)
\end{equation*}
This leads to the monotonic convergence discussed above, as follows. Suppose $0 \le t_1 \le t_2\le T$. Then the max of $J$ at time $t_1$ is greater than or equal to the max of $J$ at time $t_2$:
\begin{equation*}
    \max_{0 \le x\le \xsrc} J(x,t_1) = \max_{t_1 \le s \le T,0 \le x \le \xsrc}J(x,s) \ge
\max_{t_2 \le s \le T,0 \le x \le \xsrc}J(x,s)
= \max_{0 \le x\le \xsrc} J(x,t_2) .
\end{equation*}
Of course an analogous statement holds for the min, $\min_{0 \le x \le \xsrc}J(x,t_1)\le \min_{0 \le x \le \xsrc}J(x,t_2)$.

Intuition for the periodicity of the current can be understood based on a ``particle'' or
trajectory picture of the feedback process, where the current is defined as the
net number of trajectories passing an $x$ value per second. Note first that if
there were no feedback to the boundary at $\xsrc$, then the net current there
would vanish for any $t>0$ because of reflection: every trajectory reaching the
boundary from $x < \xsrc$ would be reversed by construction yielding zero net
flow at the boundary. With feedback, every trajectory reaching the sink $x=0$ is
placed immediately at $x=\xsrc$, the source boundary. At that boundary, the
current from non-feedback trajectories is zero because of the reflectivity
argument, and the injected number of trajectories will exactly match the number
at the sink, implying $J(x=\xsrc, t) = J(x=0, t)$.

\section{Analysis of a two-dimensional example}
\label{app:2Dcounterexample}
It is instructive to study a simple two-dimensional example in detail.
Two important features emerge: (i) there is no maximum principle because the locally defined current magnitude can increase over time away from a boundary; and (ii) for the example below, there is nevertheless a one-dimensional projection of the current which does exhibit monotonic decay.

We consider the evolution of a probability distribution $p(\vec{x},t)$ in a two-dimensional vector space $\vec{x}=(x,y)$ with $0\leq x \leq L$ and $-\infty \leq y \leq \infty$, defining an infinite rectangular strip. We take the boundaries at $x=0,L$ to be periodic, meaning
\begin{equation}
\begin{array}{lr}
p(x=0,t)=p(x=L,t), & \frac{\dee p(x=0,t)}{\dee x}=\frac{\dee p(x=L,t)}{\dee x} \; .
\end{array}
\label{eq:periodicBC}
\end{equation}
Note that this periodicity assumption is not a source-sink condition.
The probability distribution $p(x,t)$ evolves according to the continuity equation
\begin{equation}
\frac{\dee p(\vec{x},t)}{\dee t} = -\vec{\nabla}\cdot\vec{J}(\vec{x},t) \; ,
\label{eq:diffusion2D}
\end{equation}
where the Smoluchowski current $\vec{J}(\vec{x},t)$ has the usual drift and diffusion terms:
\begin{equation}
    \vec{J}(\vec{x},t)=\beta D \vec{f}(\vec{x}) p(\vec{x},t)-D\vec{\nabla}p(\vec{x},t)
    \label{eq:current2D}
\end{equation}

We consider a potential $U(\vec{x})=\frac{1}{2}k y^2-bx$ so that the force vector is $\vec{f}(\vec{x})=(-b,-ky)$. 
The constant force in $x$ is qualitatively similar to a source-sink setup, but probability can cross the boundary in both directions here.
In the $y$ direction, there is simple harmonic behavior.
The steady-state solution $p^{\infty}$ is uniform in $x$ and varies only in $y$:
\begin{equation}
p^{\infty}(\vec{x})=\frac{1}{L}\sqrt{\frac{\beta k}{2 \pi}}\exp{\frac{-\beta k y^2}{2}} \; .
\label{eq:steady_state_probability}
\end{equation}
At steady state there is a persistent current in $x$ due to periodicity but no current in the $y$ direction:
\begin{equation}
\begin{array}{lr}
J_x^{\infty}=-\beta D b p^\infty, & J_y^\infty=0 \; .
\end{array}
\label{eq:steady_state_current}
\end{equation}

Our interest is focused on the current extrema, and particularly the maximum in this case.
The maximum steady-state current magnitude, at $y=0$, is found from \eqref{eq:current2D} and \eqref{eq:steady_state_probability} to be
\begin{equation}
    |\vec{J}^{\infty}|^2_{\mathrm{max}}=\frac{\beta^3 D^2 b^2 k}{2 \pi L^2} \; .
    \label{eq:maxSSflux}
\end{equation}
To test for monotonic behavior, we employ the current magnitude resulting from an arbitrary initial condition $p^0(\vec{x},t=0)$, which is given by
\begin{equation}
    |\vec{J}^{0}|^2=D^2\bigg(\frac{\dee p^0}{\dee y}\bigg)^2+(D \beta k y p^0)^2+2\beta D^2 k y p^0 \frac{\dee p^0}{\dee y}+2\beta D^2 b p^0 \frac{\dee p^0}{\dee x}+(\beta D b p^0)^2+D^2\bigg(\frac{\dee p^0}{\dee x}\bigg)^2 .
    \label{eq:max0flux}
\end{equation}
Because the diffusion coefficient scales linearly with temperature, $D \propto \beta^{-1}$, the terms here scale as $\beta^{-2}$, $\beta^{-1}$ and $\beta^0$. From \eqref{eq:maxSSflux}, however, the steady-state flux magnitude scales as $|\vec{J}^{\infty}|^2_{\mathrm{max}}\propto\beta$, so generically, given any initial condition, the final steady-state current can be made larger than the initial current by reducing the temperature. 
This two-dimensional behavior is distinctly different from that found above in the one-dimensional case, where the current obeys a maximum principle and thus the maximum must be at the initial condition, or on the system boundary.

As a specific example, consider the initial condition of a distribution uniform in $x$ and Gaussian in $y$, $p^0(\vec{x})=\frac{1}{L}\sqrt{\frac{C}{2 \pi}}\exp{\frac{-C y^2}{2}}$, which differs from the steady distribution when $C \neq \beta k$. The initial current magnitude is then
\begin{equation}
    |\vec{J}^{0}|^2=|\vec{J}^{\infty}|^2_{\mathrm{max}}\frac{C}{\beta k}\bigg[ 1+\frac{y^2(\beta k-C)^2}{(\beta b)^2} \bigg]\exp{[-Cy^2]} \; .
    \label{eq:0fluxmag}
\end{equation}
Note that the symmetric current maxima are shifted away from $y=0$. Setting the derivative of \eqref{eq:0fluxmag} equal to zero, the maxima are located at $\ymax= \pm \sqrt{\frac{(\beta k-C)^2-C(\beta b)^2}{C(\beta k-C)^2}}$.
Inserting $\ymax$ into \eqref{eq:0fluxmag} yields
\begin{equation}
    |\vec{J}^{0}|^2_{\mathrm{max}}=|\vec{J}^{\infty}|^2_{\mathrm{max}}\bigg( \frac{(\beta k-C)^2}{\beta k (\beta b)^2 e} \bigg)\exp{\bigg[\frac{C (\beta b)^2}{(\beta k-C)^2}\bigg]}
    \label{eq:max0fluxmag}
\end{equation}
for the maximum current magnitude.

Monotonic decay of the maximum is not always observed for this system.
Over much of parameter space, $|\vec{J}^{0}|^2_{\mathrm{max}}>|\vec{J}^{\infty}|^2_{\mathrm{max}}$, but not when $(\beta b)^2>\beta k \gg C$. Defining the equilibrium root mean-square fluctuation lengths $\sigma_x=(\beta b)^{-1}$ and $\sigma_y=(\beta k)^{-1/2}$, when the width of the initial distribution is very large compared to the thermal fluctuation lengths, the ratio of initial to steady-state maximum current is $\frac{|\vec{J}^{0}|_\mathrm{{max}}}{|\vec{J}^{\infty}|_{max}}\propto\frac{\sigma_x}{\sigma_y}=(\beta^{-1/2})\frac{\sqrt{k}}{b}$. The initial current can be tuned to be less than the steady-state current by lowering the temperature, or by reducing the ratio of longitudinal ($x$) to transverse ($y$) fluctuations. This is a simple example which demonstrates that there is no maximum principle for the magnitude of the current in dimensionality exceeding one. 

Note that in this example the projected dynamics onto either the $x$ or $y$ dimension are independent because neither the potential nor the thermal noise couple $x$ and $y$. In this case, Equation \eqref{eq:diffusion2D} is fully separable with the projected currents in $x$ and $y$ each satisfying a maximum principle individually.



\bibliography{current_bound}

\begin{thebibliography}{52}%
\makeatletter
\providecommand \@ifxundefined [1]{%
 \@ifx{#1\undefined}
}%
\providecommand \@ifnum [1]{%
 \ifnum #1\expandafter \@firstoftwo
 \else \expandafter \@secondoftwo
 \fi
}%
\providecommand \@ifx [1]{%
 \ifx #1\expandafter \@firstoftwo
 \else \expandafter \@secondoftwo
 \fi
}%
\providecommand \natexlab [1]{#1}%
\providecommand \enquote  [1]{``#1''}%
\providecommand \bibnamefont  [1]{#1}%
\providecommand \bibfnamefont [1]{#1}%
\providecommand \citenamefont [1]{#1}%
\providecommand \href@noop [0]{\@secondoftwo}%
\providecommand \href [0]{\begingroup \@sanitize@url \@href}%
\providecommand \@href[1]{\@@startlink{#1}\@@href}%
\providecommand \@@href[1]{\endgroup#1\@@endlink}%
\providecommand \@sanitize@url [0]{\catcode `\\12\catcode `\$12\catcode
  `\&12\catcode `\#12\catcode `\^12\catcode `\_12\catcode `\%12\relax}%
\providecommand \@@startlink[1]{}%
\providecommand \@@endlink[0]{}%
\providecommand \url  [0]{\begingroup\@sanitize@url \@url }%
\providecommand \@url [1]{\endgroup\@href {#1}{\urlprefix }}%
\providecommand \urlprefix  [0]{URL }%
\providecommand \Eprint [0]{\href }%
\providecommand \doibase [0]{http://dx.doi.org/}%
\providecommand \selectlanguage [0]{\@gobble}%
\providecommand \bibinfo  [0]{\@secondoftwo}%
\providecommand \bibfield  [0]{\@secondoftwo}%
\providecommand \translation [1]{[#1]}%
\providecommand \BibitemOpen [0]{}%
\providecommand \bibitemStop [0]{}%
\providecommand \bibitemNoStop [0]{.\EOS\space}%
\providecommand \EOS [0]{\spacefactor3000\relax}%
\providecommand \BibitemShut  [1]{\csname bibitem#1\endcsname}%
\let\auto@bib@innerbib\@empty
\bibitem [{\citenamefont {Hopfield}(1974)}]{hopfield1974kinetic}%
  \BibitemOpen
  \bibfield  {author} {\bibinfo {author} {\bibfnamefont {J.~J.}\ \bibnamefont
  {Hopfield}},\ }\bibfield  {title} {\enquote {\bibinfo {title} {Kinetic
  proofreading: a new mechanism for reducing errors in biosynthetic processes
  requiring high specificity},}\ }\href@noop {} {\bibfield  {journal} {\bibinfo
   {journal} {Proceedings of the National Academy of Sciences}\ }\textbf
  {\bibinfo {volume} {71}},\ \bibinfo {pages} {4135--4139} (\bibinfo {year}
  {1974})}\BibitemShut {NoStop}%
\bibitem [{\citenamefont {Hill}(2004)}]{Hill2004}%
  \BibitemOpen
  \bibfield  {author} {\bibinfo {author} {\bibfnamefont {T.~L.}\ \bibnamefont
  {Hill}},\ }\href@noop {} {\emph {\bibinfo {title} {{Free Energy Transduction
  and Biochemical Cycle Kinetics}}}}\ (\bibinfo  {publisher} {Dover},\ \bibinfo
  {year} {2004})\BibitemShut {NoStop}%
\bibitem [{\citenamefont {Beard}\ and\ \citenamefont
  {Qian}(2008)}]{beard2008chemical}%
  \BibitemOpen
  \bibfield  {author} {\bibinfo {author} {\bibfnamefont {D.~A.}\ \bibnamefont
  {Beard}}\ and\ \bibinfo {author} {\bibfnamefont {H.}~\bibnamefont {Qian}},\
  }\href@noop {} {\emph {\bibinfo {title} {Chemical biophysics: quantitative
  analysis of cellular systems}}}\ (\bibinfo  {publisher} {Cambridge University
  Press},\ \bibinfo {year} {2008})\BibitemShut {NoStop}%
\bibitem [{\citenamefont {Lee}\ \emph {et~al.}(2019)\citenamefont {Lee},
  \citenamefont {Phelps}, \citenamefont {Huang}, \citenamefont {Mostofian},
  \citenamefont {Wu}, \citenamefont {Zhang}, \citenamefont {Chang},
  \citenamefont {Stork}, \citenamefont {Gray}, \citenamefont {Zuckerman} \emph
  {et~al.}}]{lee2019high}%
  \BibitemOpen
  \bibfield  {author} {\bibinfo {author} {\bibfnamefont {Y.}~\bibnamefont
  {Lee}}, \bibinfo {author} {\bibfnamefont {C.}~\bibnamefont {Phelps}},
  \bibinfo {author} {\bibfnamefont {T.}~\bibnamefont {Huang}}, \bibinfo
  {author} {\bibfnamefont {B.}~\bibnamefont {Mostofian}}, \bibinfo {author}
  {\bibfnamefont {L.}~\bibnamefont {Wu}}, \bibinfo {author} {\bibfnamefont
  {Y.}~\bibnamefont {Zhang}}, \bibinfo {author} {\bibfnamefont {Y.~H.}\
  \bibnamefont {Chang}}, \bibinfo {author} {\bibfnamefont {P.~J.}\ \bibnamefont
  {Stork}}, \bibinfo {author} {\bibfnamefont {J.~W.}\ \bibnamefont {Gray}},
  \bibinfo {author} {\bibfnamefont {D.~M.}\ \bibnamefont {Zuckerman}},  \emph
  {et~al.},\ }\bibfield  {title} {\enquote {\bibinfo {title} {High-throughput
  single-particle tracking reveals nested membrane nanodomain organization that
  dictates ras diffusion and trafficking},}\ }\href@noop {} {\bibfield
  {journal} {\bibinfo  {journal} {bioRxiv}\ ,\ \bibinfo {pages} {552075}}
  (\bibinfo {year} {2019})}\BibitemShut {NoStop}%
\bibitem [{\citenamefont {Jarzynski}(1997)}]{jarzynski1997nonequilibrium}%
  \BibitemOpen
  \bibfield  {author} {\bibinfo {author} {\bibfnamefont {C.}~\bibnamefont
  {Jarzynski}},\ }\bibfield  {title} {\enquote {\bibinfo {title}
  {Nonequilibrium equality for free energy differences},}\ }\href@noop {}
  {\bibfield  {journal} {\bibinfo  {journal} {Physical Review Letters}\
  }\textbf {\bibinfo {volume} {78}},\ \bibinfo {pages} {2690} (\bibinfo {year}
  {1997})}\BibitemShut {NoStop}%
\bibitem [{\citenamefont {Crooks}(1999)}]{crooks1999entropy}%
  \BibitemOpen
  \bibfield  {author} {\bibinfo {author} {\bibfnamefont {G.~E.}\ \bibnamefont
  {Crooks}},\ }\bibfield  {title} {\enquote {\bibinfo {title} {Entropy
  production fluctuation theorem and the nonequilibrium work relation for free
  energy differences},}\ }\href@noop {} {\bibfield  {journal} {\bibinfo
  {journal} {Physical Review E}\ }\textbf {\bibinfo {volume} {60}},\ \bibinfo
  {pages} {2721} (\bibinfo {year} {1999})}\BibitemShut {NoStop}%
\bibitem [{\citenamefont {Seifert}(2012)}]{seifert2012stochastic}%
  \BibitemOpen
  \bibfield  {author} {\bibinfo {author} {\bibfnamefont {U.}~\bibnamefont
  {Seifert}},\ }\bibfield  {title} {\enquote {\bibinfo {title} {Stochastic
  thermodynamics, fluctuation theorems and molecular machines},}\ }\href@noop
  {} {\bibfield  {journal} {\bibinfo  {journal} {Reports on Progress in
  Physics}\ }\textbf {\bibinfo {volume} {75}},\ \bibinfo {pages} {126001}
  (\bibinfo {year} {2012})}\BibitemShut {NoStop}%
\bibitem [{\citenamefont {Zhang}, \citenamefont {Jasnow},\ and\ \citenamefont
  {Zuckerman}(2010)}]{zhang2010weighted}%
  \BibitemOpen
  \bibfield  {author} {\bibinfo {author} {\bibfnamefont {B.~W.}\ \bibnamefont
  {Zhang}}, \bibinfo {author} {\bibfnamefont {D.}~\bibnamefont {Jasnow}}, \
  and\ \bibinfo {author} {\bibfnamefont {D.~M.}\ \bibnamefont {Zuckerman}},\
  }\bibfield  {title} {\enquote {\bibinfo {title} {The “weighted ensemble”
  path sampling method is statistically exact for a broad class of stochastic
  processes and binning procedures},}\ }\href@noop {} {\bibfield  {journal}
  {\bibinfo  {journal} {The Journal of Chemical Physics}\ }\textbf {\bibinfo
  {volume} {132}},\ \bibinfo {pages} {054107} (\bibinfo {year}
  {2010})}\BibitemShut {NoStop}%
\bibitem [{\citenamefont {Dickson}\ and\ \citenamefont
  {Dinner}(2010)}]{Dickson2010a}%
  \BibitemOpen
  \bibfield  {author} {\bibinfo {author} {\bibfnamefont {A.}~\bibnamefont
  {Dickson}}\ and\ \bibinfo {author} {\bibfnamefont {A.~R.}\ \bibnamefont
  {Dinner}},\ }\bibfield  {title} {\enquote {\bibinfo {title} {{Enhanced
  sampling of nonequilibrium steady states.}}}\ }\href {\doibase
  10.1146/annurev.physchem.012809.103433} {\bibfield  {journal} {\bibinfo
  {journal} {Annual Review of Physical Chemistry}\ }\textbf {\bibinfo {volume}
  {61}},\ \bibinfo {pages} {441--459} (\bibinfo {year} {2010})}\BibitemShut
  {NoStop}%
\bibitem [{\citenamefont {Ytreberg}\ and\ \citenamefont
  {Zuckerman}(2004)}]{ytreberg2004single}%
  \BibitemOpen
  \bibfield  {author} {\bibinfo {author} {\bibfnamefont {F.~M.}\ \bibnamefont
  {Ytreberg}}\ and\ \bibinfo {author} {\bibfnamefont {D.~M.}\ \bibnamefont
  {Zuckerman}},\ }\bibfield  {title} {\enquote {\bibinfo {title}
  {Single-ensemble nonequilibrium path-sampling estimates of free energy
  differences},}\ }\href@noop {} {\bibfield  {journal} {\bibinfo  {journal}
  {The Journal of Chemical Physics}\ }\textbf {\bibinfo {volume} {120}},\
  \bibinfo {pages} {10876--10879} (\bibinfo {year} {2004})}\BibitemShut
  {NoStop}%
\bibitem [{\citenamefont {Bello-Rivas}\ and\ \citenamefont
  {Elber}(2015)}]{bello2015exact}%
  \BibitemOpen
  \bibfield  {author} {\bibinfo {author} {\bibfnamefont {J.~M.}\ \bibnamefont
  {Bello-Rivas}}\ and\ \bibinfo {author} {\bibfnamefont {R.}~\bibnamefont
  {Elber}},\ }\bibfield  {title} {\enquote {\bibinfo {title} {Exact
  milestoning},}\ }\href@noop {} {\bibfield  {journal} {\bibinfo  {journal}
  {The Journal of Chemical Physics}\ }\textbf {\bibinfo {volume} {142}},\
  \bibinfo {pages} {03B602\_1} (\bibinfo {year} {2015})}\BibitemShut {NoStop}%
\bibitem [{\citenamefont {Nilmeier}\ \emph {et~al.}(2011)\citenamefont
  {Nilmeier}, \citenamefont {Crooks}, \citenamefont {Minh},\ and\ \citenamefont
  {Chodera}}]{nilmeier2011nonequilibrium}%
  \BibitemOpen
  \bibfield  {author} {\bibinfo {author} {\bibfnamefont {J.~P.}\ \bibnamefont
  {Nilmeier}}, \bibinfo {author} {\bibfnamefont {G.~E.}\ \bibnamefont
  {Crooks}}, \bibinfo {author} {\bibfnamefont {D.~D.}\ \bibnamefont {Minh}}, \
  and\ \bibinfo {author} {\bibfnamefont {J.~D.}\ \bibnamefont {Chodera}},\
  }\bibfield  {title} {\enquote {\bibinfo {title} {Nonequilibrium candidate
  {Monte Carlo} is an efficient tool for equilibrium simulation},}\ }\href@noop
  {} {\bibfield  {journal} {\bibinfo  {journal} {Proceedings of the National
  Academy of Sciences}\ } (\bibinfo {year} {2011})}\BibitemShut {NoStop}%
\bibitem [{\citenamefont {Chapman}(1928)}]{chapman1928brownian}%
  \BibitemOpen
  \bibfield  {author} {\bibinfo {author} {\bibfnamefont {S.}~\bibnamefont
  {Chapman}},\ }\bibfield  {title} {\enquote {\bibinfo {title} {On the brownian
  displacements and thermal diffusion of grains suspended in a non-uniform
  fluid},}\ }\href@noop {} {\bibfield  {journal} {\bibinfo  {journal}
  {Proceedings of the Royal Society of London. Series A, Containing Papers of a
  Mathematical and Physical Character}\ }\textbf {\bibinfo {volume} {119}},\
  \bibinfo {pages} {34--54} (\bibinfo {year} {1928})}\BibitemShut {NoStop}%
\bibitem [{\citenamefont {Kolmogoroff}(1931)}]{kolmogoroff1931analytical}%
  \BibitemOpen
  \bibfield  {author} {\bibinfo {author} {\bibfnamefont {A.}~\bibnamefont
  {Kolmogoroff}},\ }\bibfield  {title} {\enquote {\bibinfo {title} {On
  analytical methods in the theory of probability},}\ }\href@noop {} {\bibfield
   {journal} {\bibinfo  {journal} {Mathematische Annalen}\ }\textbf {\bibinfo
  {volume} {104}},\ \bibinfo {pages} {415--458} (\bibinfo {year}
  {1931})}\BibitemShut {NoStop}%
\bibitem [{\citenamefont {Van~Kampen}(1992)}]{van1992stochastic}%
  \BibitemOpen
  \bibfield  {author} {\bibinfo {author} {\bibfnamefont {N.~G.}\ \bibnamefont
  {Van~Kampen}},\ }\href@noop {} {\emph {\bibinfo {title} {Stochastic processes
  in physics and chemistry}}},\ Vol.~\bibinfo {volume} {1}\ (\bibinfo
  {publisher} {Elsevier},\ \bibinfo {year} {1992})\BibitemShut {NoStop}%
\bibitem [{\citenamefont {Risken}\ and\ \citenamefont
  {Frank}(1996)}]{Risken1996}%
  \BibitemOpen
  \bibfield  {author} {\bibinfo {author} {\bibfnamefont {H.}~\bibnamefont
  {Risken}}\ and\ \bibinfo {author} {\bibfnamefont {T.}~\bibnamefont {Frank}},\
  }\href@noop {} {\emph {\bibinfo {title} {The Fokker-Planck Equation: Methods
  of Solution and Applications}}},\ Vol.~\bibinfo {volume} {18}\ (\bibinfo
  {publisher} {Springer Science \& Business Media},\ \bibinfo {year}
  {1996})\BibitemShut {NoStop}%
\bibitem [{\citenamefont {Gardiner}(2009)}]{gardiner2009stochastic}%
  \BibitemOpen
  \bibfield  {author} {\bibinfo {author} {\bibfnamefont {C.}~\bibnamefont
  {Gardiner}},\ }\href@noop {} {\emph {\bibinfo {title} {Stochastic
  methods}}},\ Vol.~\bibinfo {volume} {4}\ (\bibinfo  {publisher} {Springer
  Berlin},\ \bibinfo {year} {2009})\BibitemShut {NoStop}%
\bibitem [{\citenamefont {Berezhkovskii}\ \emph {et~al.}(2014)\citenamefont
  {Berezhkovskii}, \citenamefont {Szabo}, \citenamefont {Greives},\ and\
  \citenamefont {Zhou}}]{berezhkovskii2014multidimensional}%
  \BibitemOpen
  \bibfield  {author} {\bibinfo {author} {\bibfnamefont {A.~M.}\ \bibnamefont
  {Berezhkovskii}}, \bibinfo {author} {\bibfnamefont {A.}~\bibnamefont
  {Szabo}}, \bibinfo {author} {\bibfnamefont {N.}~\bibnamefont {Greives}}, \
  and\ \bibinfo {author} {\bibfnamefont {H.-X.}\ \bibnamefont {Zhou}},\
  }\bibfield  {title} {\enquote {\bibinfo {title} {Multidimensional reaction
  rate theory with anisotropic diffusion},}\ }\href@noop {} {\bibfield
  {journal} {\bibinfo  {journal} {The Journal of Chemical Physics}\ }\textbf
  {\bibinfo {volume} {141}},\ \bibinfo {pages} {11B616\_1} (\bibinfo {year}
  {2014})}\BibitemShut {NoStop}%
\bibitem [{\citenamefont {Ghysels}\ \emph {et~al.}(2017)\citenamefont
  {Ghysels}, \citenamefont {Venable}, \citenamefont {Pastor},\ and\
  \citenamefont {Hummer}}]{ghysels2017position}%
  \BibitemOpen
  \bibfield  {author} {\bibinfo {author} {\bibfnamefont {A.}~\bibnamefont
  {Ghysels}}, \bibinfo {author} {\bibfnamefont {R.~M.}\ \bibnamefont
  {Venable}}, \bibinfo {author} {\bibfnamefont {R.~W.}\ \bibnamefont {Pastor}},
  \ and\ \bibinfo {author} {\bibfnamefont {G.}~\bibnamefont {Hummer}},\
  }\bibfield  {title} {\enquote {\bibinfo {title} {Position-dependent diffusion
  tensors in anisotropic media from simulation: oxygen transport in and through
  membranes},}\ }\href@noop {} {\bibfield  {journal} {\bibinfo  {journal}
  {Journal of Chemical Theory and Computation}\ }\textbf {\bibinfo {volume}
  {13}},\ \bibinfo {pages} {2962--2976} (\bibinfo {year} {2017})}\BibitemShut
  {NoStop}%
\bibitem [{\citenamefont {Grafke}\ and\ \citenamefont
  {Vanden-Eijnden}(2019)}]{grafke2018numerical}%
  \BibitemOpen
  \bibfield  {author} {\bibinfo {author} {\bibfnamefont {T.}~\bibnamefont
  {Grafke}}\ and\ \bibinfo {author} {\bibfnamefont {E.}~\bibnamefont
  {Vanden-Eijnden}},\ }\bibfield  {title} {\enquote {\bibinfo {title}
  {Numerical computation of rare events via large deviation theory},}\
  }\href@noop {} {\bibfield  {journal} {\bibinfo  {journal} {Chaos: An
  Interdisciplinary Journal of Nonlinear Science}\ }\textbf {\bibinfo {volume}
  {29}},\ \bibinfo {pages} {063118} (\bibinfo {year} {2019})}\BibitemShut
  {NoStop}%
\bibitem [{\citenamefont {Polotto}\ \emph {et~al.}(2018)\citenamefont
  {Polotto}, \citenamefont {Drigo~Filho}, \citenamefont {Chahine},\ and\
  \citenamefont {de~Oliveira}}]{polotto2018supersymmetric}%
  \BibitemOpen
  \bibfield  {author} {\bibinfo {author} {\bibfnamefont {F.}~\bibnamefont
  {Polotto}}, \bibinfo {author} {\bibfnamefont {E.}~\bibnamefont
  {Drigo~Filho}}, \bibinfo {author} {\bibfnamefont {J.}~\bibnamefont
  {Chahine}}, \ and\ \bibinfo {author} {\bibfnamefont {R.~J.}\ \bibnamefont
  {de~Oliveira}},\ }\bibfield  {title} {\enquote {\bibinfo {title}
  {Supersymmetric quantum mechanics method for the fokker--planck equation with
  applications to protein folding dynamics},}\ }\href@noop {} {\bibfield
  {journal} {\bibinfo  {journal} {Physica A: Statistical Mechanics and its
  Applications}\ }\textbf {\bibinfo {volume} {493}},\ \bibinfo {pages}
  {286--300} (\bibinfo {year} {2018})}\BibitemShut {NoStop}%
\bibitem [{\citenamefont {Cossio}, \citenamefont {Hummer},\ and\ \citenamefont
  {Szabo}(2018)}]{cossio2018transition}%
  \BibitemOpen
  \bibfield  {author} {\bibinfo {author} {\bibfnamefont {P.}~\bibnamefont
  {Cossio}}, \bibinfo {author} {\bibfnamefont {G.}~\bibnamefont {Hummer}}, \
  and\ \bibinfo {author} {\bibfnamefont {A.}~\bibnamefont {Szabo}},\ }\bibfield
   {title} {\enquote {\bibinfo {title} {Transition paths in single-molecule
  force spectroscopy},}\ }\href@noop {} {\bibfield  {journal} {\bibinfo
  {journal} {The Journal of Chemical Physics}\ }\textbf {\bibinfo {volume}
  {148}},\ \bibinfo {pages} {123309} (\bibinfo {year} {2018})}\BibitemShut
  {NoStop}%
\bibitem [{\citenamefont {del Razo}, \citenamefont {Qian},\ and\ \citenamefont
  {No{\'e}}(2018)}]{del2018grand}%
  \BibitemOpen
  \bibfield  {author} {\bibinfo {author} {\bibfnamefont {M.~J.}\ \bibnamefont
  {del Razo}}, \bibinfo {author} {\bibfnamefont {H.}~\bibnamefont {Qian}}, \
  and\ \bibinfo {author} {\bibfnamefont {F.}~\bibnamefont {No{\'e}}},\
  }\bibfield  {title} {\enquote {\bibinfo {title} {Grand canonical
  diffusion-influenced reactions: A stochastic theory with applications to
  multiscale reaction-diffusion simulations},}\ }\href@noop {} {\bibfield
  {journal} {\bibinfo  {journal} {The Journal of Chemical Physics}\ }\textbf
  {\bibinfo {volume} {149}},\ \bibinfo {pages} {044102} (\bibinfo {year}
  {2018})}\BibitemShut {NoStop}%
\bibitem [{\citenamefont {Berezhkovskii}\ and\ \citenamefont
  {Bezrukov}(2018)}]{berezhkovskii2018mapping}%
  \BibitemOpen
  \bibfield  {author} {\bibinfo {author} {\bibfnamefont {A.~M.}\ \bibnamefont
  {Berezhkovskii}}\ and\ \bibinfo {author} {\bibfnamefont {S.~M.}\ \bibnamefont
  {Bezrukov}},\ }\bibfield  {title} {\enquote {\bibinfo {title} {Mapping
  intrachannel diffusive dynamics of interacting molecules onto a two-site
  model: Crossover in flux concentration dependence},}\ }\href@noop {}
  {\bibfield  {journal} {\bibinfo  {journal} {The Journal of Physical Chemistry
  B}\ }\textbf {\bibinfo {volume} {122}},\ \bibinfo {pages} {10996--11001}
  (\bibinfo {year} {2018})}\BibitemShut {NoStop}%
\bibitem [{\citenamefont {Singhal}, \citenamefont {Snow},\ and\ \citenamefont
  {Pande}(2004)}]{singhal2004using}%
  \BibitemOpen
  \bibfield  {author} {\bibinfo {author} {\bibfnamefont {N.}~\bibnamefont
  {Singhal}}, \bibinfo {author} {\bibfnamefont {C.~D.}\ \bibnamefont {Snow}}, \
  and\ \bibinfo {author} {\bibfnamefont {V.~S.}\ \bibnamefont {Pande}},\
  }\bibfield  {title} {\enquote {\bibinfo {title} {Using path sampling to build
  better {Markovian} state models: predicting the folding rate and mechanism of
  a tryptophan zipper beta hairpin},}\ }\href@noop {} {\bibfield  {journal}
  {\bibinfo  {journal} {The Journal of Chemical Physics}\ }\textbf {\bibinfo
  {volume} {121}},\ \bibinfo {pages} {415--425} (\bibinfo {year}
  {2004})}\BibitemShut {NoStop}%
\bibitem [{\citenamefont {No{\'e}}\ \emph {et~al.}(2007)\citenamefont
  {No{\'e}}, \citenamefont {Horenko}, \citenamefont {Sch{\"u}tte},\ and\
  \citenamefont {Smith}}]{noe2007hierarchical}%
  \BibitemOpen
  \bibfield  {author} {\bibinfo {author} {\bibfnamefont {F.}~\bibnamefont
  {No{\'e}}}, \bibinfo {author} {\bibfnamefont {I.}~\bibnamefont {Horenko}},
  \bibinfo {author} {\bibfnamefont {C.}~\bibnamefont {Sch{\"u}tte}}, \ and\
  \bibinfo {author} {\bibfnamefont {J.~C.}\ \bibnamefont {Smith}},\ }\bibfield
  {title} {\enquote {\bibinfo {title} {Hierarchical analysis of conformational
  dynamics in biomolecules: transition networks of metastable states},}\
  }\href@noop {} {\bibfield  {journal} {\bibinfo  {journal} {The Journal of
  Chemical Physics}\ }\textbf {\bibinfo {volume} {126}},\ \bibinfo {pages}
  {04B617} (\bibinfo {year} {2007})}\BibitemShut {NoStop}%
\bibitem [{\citenamefont {Voelz}\ \emph {et~al.}(2010)\citenamefont {Voelz},
  \citenamefont {Bowman}, \citenamefont {Beauchamp},\ and\ \citenamefont
  {Pande}}]{voelz2010molecular}%
  \BibitemOpen
  \bibfield  {author} {\bibinfo {author} {\bibfnamefont {V.~A.}\ \bibnamefont
  {Voelz}}, \bibinfo {author} {\bibfnamefont {G.~R.}\ \bibnamefont {Bowman}},
  \bibinfo {author} {\bibfnamefont {K.}~\bibnamefont {Beauchamp}}, \ and\
  \bibinfo {author} {\bibfnamefont {V.~S.}\ \bibnamefont {Pande}},\ }\bibfield
  {title} {\enquote {\bibinfo {title} {Molecular simulation of ab initio
  protein folding for a millisecond folder {NTL9} (1- 39)},}\ }\href@noop {}
  {\bibfield  {journal} {\bibinfo  {journal} {Journal of the American Chemical
  Society}\ }\textbf {\bibinfo {volume} {132}},\ \bibinfo {pages} {1526--1528}
  (\bibinfo {year} {2010})}\BibitemShut {NoStop}%
\bibitem [{\citenamefont {Chodera}\ and\ \citenamefont
  {No{\'e}}(2014)}]{chodera2014markov}%
  \BibitemOpen
  \bibfield  {author} {\bibinfo {author} {\bibfnamefont {J.~D.}\ \bibnamefont
  {Chodera}}\ and\ \bibinfo {author} {\bibfnamefont {F.}~\bibnamefont
  {No{\'e}}},\ }\bibfield  {title} {\enquote {\bibinfo {title} {Markov state
  models of biomolecular conformational dynamics},}\ }\href@noop {} {\bibfield
  {journal} {\bibinfo  {journal} {Current Opinion in Structural Biology}\
  }\textbf {\bibinfo {volume} {25}},\ \bibinfo {pages} {135--144} (\bibinfo
  {year} {2014})}\BibitemShut {NoStop}%
\bibitem [{\citenamefont {Makarov}(2015)}]{makarov2015single}%
  \BibitemOpen
  \bibfield  {author} {\bibinfo {author} {\bibfnamefont {D.~E.}\ \bibnamefont
  {Makarov}},\ }\href@noop {} {\emph {\bibinfo {title} {Single Molecule
  Science: Physical Principles and Models}}}\ (\bibinfo  {publisher} {CRC
  Press},\ \bibinfo {year} {2015})\BibitemShut {NoStop}%
\bibitem [{\citenamefont {Huber}\ and\ \citenamefont
  {Kim}(1996)}]{huber1996weighted}%
  \BibitemOpen
  \bibfield  {author} {\bibinfo {author} {\bibfnamefont {G.~A.}\ \bibnamefont
  {Huber}}\ and\ \bibinfo {author} {\bibfnamefont {S.}~\bibnamefont {Kim}},\
  }\bibfield  {title} {\enquote {\bibinfo {title} {Weighted-ensemble brownian
  dynamics simulations for protein association reactions.}}\ }\href@noop {}
  {\bibfield  {journal} {\bibinfo  {journal} {Biophysical Journal}\ }\textbf
  {\bibinfo {volume} {70}},\ \bibinfo {pages} {97} (\bibinfo {year}
  {1996})}\BibitemShut {NoStop}%
\bibitem [{\citenamefont {Suarez}\ \emph {et~al.}(2014)\citenamefont {Suarez},
  \citenamefont {Lettieri}, \citenamefont {Zwier}, \citenamefont {Stringer},
  \citenamefont {Subramanian}, \citenamefont {Chong},\ and\ \citenamefont
  {Zuckerman}}]{suarez2014simultaneous}%
  \BibitemOpen
  \bibfield  {author} {\bibinfo {author} {\bibfnamefont {E.}~\bibnamefont
  {Suarez}}, \bibinfo {author} {\bibfnamefont {S.}~\bibnamefont {Lettieri}},
  \bibinfo {author} {\bibfnamefont {M.~C.}\ \bibnamefont {Zwier}}, \bibinfo
  {author} {\bibfnamefont {C.~A.}\ \bibnamefont {Stringer}}, \bibinfo {author}
  {\bibfnamefont {S.~R.}\ \bibnamefont {Subramanian}}, \bibinfo {author}
  {\bibfnamefont {L.~T.}\ \bibnamefont {Chong}}, \ and\ \bibinfo {author}
  {\bibfnamefont {D.~M.}\ \bibnamefont {Zuckerman}},\ }\bibfield  {title}
  {\enquote {\bibinfo {title} {Simultaneous computation of dynamical and
  equilibrium information using a weighted ensemble of trajectories},}\
  }\href@noop {} {\bibfield  {journal} {\bibinfo  {journal} {Journal of
  Chemical Theory and Computation}\ }\textbf {\bibinfo {volume} {10}},\
  \bibinfo {pages} {2658--2667} (\bibinfo {year} {2014})}\BibitemShut {NoStop}%
\bibitem [{\citenamefont {Adhikari}\ \emph {et~al.}(2019)\citenamefont
  {Adhikari}, \citenamefont {Mostofian}, \citenamefont {Copperman},
  \citenamefont {Subramanian}, \citenamefont {Petersen},\ and\ \citenamefont
  {Zuckerman}}]{adhikari2018computational}%
  \BibitemOpen
  \bibfield  {author} {\bibinfo {author} {\bibfnamefont {U.}~\bibnamefont
  {Adhikari}}, \bibinfo {author} {\bibfnamefont {B.}~\bibnamefont {Mostofian}},
  \bibinfo {author} {\bibfnamefont {J.}~\bibnamefont {Copperman}}, \bibinfo
  {author} {\bibfnamefont {S.~R.}\ \bibnamefont {Subramanian}}, \bibinfo
  {author} {\bibfnamefont {A.~A.}\ \bibnamefont {Petersen}}, \ and\ \bibinfo
  {author} {\bibfnamefont {D.~M.}\ \bibnamefont {Zuckerman}},\ }\bibfield
  {title} {\enquote {\bibinfo {title} {Computational estimation of microsecond
  to second atomistic folding times},}\ }\href@noop {} {\bibfield  {journal}
  {\bibinfo  {journal} {Journal of the American Chemical Society}\ }\textbf
  {\bibinfo {volume} {141}},\ \bibinfo {pages} {6519--6526} (\bibinfo {year}
  {2019})}\BibitemShut {NoStop}%
\bibitem [{\citenamefont {Bhatt}, \citenamefont {Zhang},\ and\ \citenamefont
  {Zuckerman}(2010)}]{bhatt2010steady}%
  \BibitemOpen
  \bibfield  {author} {\bibinfo {author} {\bibfnamefont {D.}~\bibnamefont
  {Bhatt}}, \bibinfo {author} {\bibfnamefont {B.~W.}\ \bibnamefont {Zhang}}, \
  and\ \bibinfo {author} {\bibfnamefont {D.~M.}\ \bibnamefont {Zuckerman}},\
  }\bibfield  {title} {\enquote {\bibinfo {title} {Steady-state simulations
  using weighted ensemble path sampling},}\ }\href@noop {} {\bibfield
  {journal} {\bibinfo  {journal} {The Journal of Chemical Physics}\ }\textbf
  {\bibinfo {volume} {133}},\ \bibinfo {pages} {014110} (\bibinfo {year}
  {2010})}\BibitemShut {NoStop}%
\bibitem [{\citenamefont {H{\"a}nggi}, \citenamefont {Talkner},\ and\
  \citenamefont {Borkovec}(1990)}]{hanggi1990reaction}%
  \BibitemOpen
  \bibfield  {author} {\bibinfo {author} {\bibfnamefont {P.}~\bibnamefont
  {H{\"a}nggi}}, \bibinfo {author} {\bibfnamefont {P.}~\bibnamefont {Talkner}},
  \ and\ \bibinfo {author} {\bibfnamefont {M.}~\bibnamefont {Borkovec}},\
  }\bibfield  {title} {\enquote {\bibinfo {title} {Reaction-rate theory: fifty
  years after kramers},}\ }\href@noop {} {\bibfield  {journal} {\bibinfo
  {journal} {Reviews of Modern Physics}\ }\textbf {\bibinfo {volume} {62}},\
  \bibinfo {pages} {251} (\bibinfo {year} {1990})}\BibitemShut {NoStop}%
\bibitem [{\citenamefont {Warmflash}, \citenamefont {Bhimalapuram},\ and\
  \citenamefont {Dinner}(2007)}]{warmflash2007umbrella}%
  \BibitemOpen
  \bibfield  {author} {\bibinfo {author} {\bibfnamefont {A.}~\bibnamefont
  {Warmflash}}, \bibinfo {author} {\bibfnamefont {P.}~\bibnamefont
  {Bhimalapuram}}, \ and\ \bibinfo {author} {\bibfnamefont {A.~R.}\
  \bibnamefont {Dinner}},\ }\bibfield  {title} {\enquote {\bibinfo {title}
  {Umbrella sampling for nonequilibrium processes},}\ }\href@noop {} {\bibfield
   {journal} {\bibinfo  {journal} {The Journal of Chemical Physics}\ }\textbf
  {\bibinfo {volume} {127}},\ \bibinfo {pages} {114109} (\bibinfo {year}
  {2007})}\BibitemShut {NoStop}%
\bibitem [{\citenamefont {Dickson}, \citenamefont {Warmflash},\ and\
  \citenamefont {Dinner}(2009)}]{dickson2009nonequilibrium}%
  \BibitemOpen
  \bibfield  {author} {\bibinfo {author} {\bibfnamefont {A.}~\bibnamefont
  {Dickson}}, \bibinfo {author} {\bibfnamefont {A.}~\bibnamefont {Warmflash}},
  \ and\ \bibinfo {author} {\bibfnamefont {A.~R.}\ \bibnamefont {Dinner}},\
  }\bibfield  {title} {\enquote {\bibinfo {title} {Nonequilibrium umbrella
  sampling in spaces of many order parameters},}\ }\href@noop {} {\bibfield
  {journal} {\bibinfo  {journal} {The Journal of Chemical Physics}\ }\textbf
  {\bibinfo {volume} {130}},\ \bibinfo {pages} {02B605} (\bibinfo {year}
  {2009})}\BibitemShut {NoStop}%
\bibitem [{\citenamefont {Aristoff}\ and\ \citenamefont
  {Zuckerman}(2018)}]{aristoff2018optimizing}%
  \BibitemOpen
  \bibfield  {author} {\bibinfo {author} {\bibfnamefont {D.}~\bibnamefont
  {Aristoff}}\ and\ \bibinfo {author} {\bibfnamefont {D.~M.}\ \bibnamefont
  {Zuckerman}},\ }\bibfield  {title} {\enquote {\bibinfo {title} {Optimizing
  weighted ensemble sampling of steady states},}\ }\href@noop {} {\bibfield
  {journal} {\bibinfo  {journal} {arXiv preprint arXiv:1806.00860}\ } (\bibinfo
  {year} {2018})}\BibitemShut {NoStop}%
\bibitem [{\citenamefont {Zuckerman}(2015)}]{zuckerman2015statistical}%
  \BibitemOpen
  \bibfield  {author} {\bibinfo {author} {\bibfnamefont {D.~M.}\ \bibnamefont
  {Zuckerman}},\ }\href@noop {} {\enquote {\bibinfo {title} {Statistical
  biophysics blog:``proof'' of the hill relation between probability flux and
  mean first-passage time},}\ }\bibinfo {howpublished}
  {\url{http://statisticalbiophysicsblog.org/?p=8}} (\bibinfo {year}
  {2015})\BibitemShut {NoStop}%
\bibitem [{\citenamefont {Metzler}\ and\ \citenamefont
  {Klafter}(2000)}]{metzler2000random}%
  \BibitemOpen
  \bibfield  {author} {\bibinfo {author} {\bibfnamefont {R.}~\bibnamefont
  {Metzler}}\ and\ \bibinfo {author} {\bibfnamefont {J.}~\bibnamefont
  {Klafter}},\ }\bibfield  {title} {\enquote {\bibinfo {title} {The random
  walk's guide to anomalous diffusion: a fractional dynamics approach},}\
  }\href@noop {} {\bibfield  {journal} {\bibinfo  {journal} {Physics reports}\
  }\textbf {\bibinfo {volume} {339}},\ \bibinfo {pages} {1--77} (\bibinfo
  {year} {2000})}\BibitemShut {NoStop}%
\bibitem [{\citenamefont {Protter}\ and\ \citenamefont
  {Weinberger}(2012)}]{protter2012maximum}%
  \BibitemOpen
  \bibfield  {author} {\bibinfo {author} {\bibfnamefont {M.~H.}\ \bibnamefont
  {Protter}}\ and\ \bibinfo {author} {\bibfnamefont {H.~F.}\ \bibnamefont
  {Weinberger}},\ }\href@noop {} {\emph {\bibinfo {title} {Maximum principles
  in differential equations}}}\ (\bibinfo  {publisher} {Springer Science \&
  Business Media},\ \bibinfo {year} {2012})\BibitemShut {NoStop}%
\bibitem [{\citenamefont {Guyer}, \citenamefont {Wheeler},\ and\ \citenamefont
  {Warren}(2009)}]{guyer2009fipy}%
  \BibitemOpen
  \bibfield  {author} {\bibinfo {author} {\bibfnamefont {J.~E.}\ \bibnamefont
  {Guyer}}, \bibinfo {author} {\bibfnamefont {D.}~\bibnamefont {Wheeler}}, \
  and\ \bibinfo {author} {\bibfnamefont {J.~A.}\ \bibnamefont {Warren}},\
  }\bibfield  {title} {\enquote {\bibinfo {title} {{FiPy}: Partial differential
  equations with python},}\ }\href@noop {} {\bibfield  {journal} {\bibinfo
  {journal} {Computing in Science \& Engineering}\ }\textbf {\bibinfo {volume}
  {11}} (\bibinfo {year} {2009})}\BibitemShut {NoStop}%
\bibitem [{\citenamefont {Berezhkovskii}\ and\ \citenamefont
  {Szabo}(2013)}]{berezhkovskii2013diffusion}%
  \BibitemOpen
  \bibfield  {author} {\bibinfo {author} {\bibfnamefont {A.~M.}\ \bibnamefont
  {Berezhkovskii}}\ and\ \bibinfo {author} {\bibfnamefont {A.}~\bibnamefont
  {Szabo}},\ }\bibfield  {title} {\enquote {\bibinfo {title} {Diffusion along
  the splitting/commitment probability reaction coordinate},}\ }\href@noop {}
  {\bibfield  {journal} {\bibinfo  {journal} {The Journal of Physical Chemistry
  B}\ }\textbf {\bibinfo {volume} {117}},\ \bibinfo {pages} {13115--13119}
  (\bibinfo {year} {2013})}\BibitemShut {NoStop}%
\bibitem [{\citenamefont {Lu}\ and\ \citenamefont
  {Vanden-Eijnden}(2014)}]{lu2014exact}%
  \BibitemOpen
  \bibfield  {author} {\bibinfo {author} {\bibfnamefont {J.}~\bibnamefont
  {Lu}}\ and\ \bibinfo {author} {\bibfnamefont {E.}~\bibnamefont
  {Vanden-Eijnden}},\ }\bibfield  {title} {\enquote {\bibinfo {title} {Exact
  dynamical coarse-graining without time-scale separation},}\ }\href@noop {}
  {\bibfield  {journal} {\bibinfo  {journal} {The Journal of Chemical Physics}\
  }\textbf {\bibinfo {volume} {141}},\ \bibinfo {pages} {07B619\_1} (\bibinfo
  {year} {2014})}\BibitemShut {NoStop}%
\bibitem [{\citenamefont {Vanden-Eijnden}(2003)}]{vanden2003fast}%
  \BibitemOpen
  \bibfield  {author} {\bibinfo {author} {\bibfnamefont {E.}~\bibnamefont
  {Vanden-Eijnden}},\ }\bibfield  {title} {\enquote {\bibinfo {title} {Fast
  communications: Numerical techniques for multi-scale dynamical systems with
  stochastic effects},}\ }\href@noop {} {\bibfield  {journal} {\bibinfo
  {journal} {Communications in Mathematical Sciences}\ }\textbf {\bibinfo
  {volume} {1}},\ \bibinfo {pages} {385--391} (\bibinfo {year}
  {2003})}\BibitemShut {NoStop}%
\bibitem [{\citenamefont {Hartmann}(2007)}]{hartmann2007model}%
  \BibitemOpen
  \bibfield  {author} {\bibinfo {author} {\bibfnamefont {C.}~\bibnamefont
  {Hartmann}},\ }\emph {\bibinfo {title} {Model reduction in classical
  molecular dynamics}},\ \href@noop {} {Ph.D. thesis},\ \bibinfo  {school}
  {Citeseer} (\bibinfo {year} {2007})\BibitemShut {NoStop}%
\bibitem [{\citenamefont {Legoll}\ and\ \citenamefont
  {Lelievre}(2010)}]{legoll2010effective}%
  \BibitemOpen
  \bibfield  {author} {\bibinfo {author} {\bibfnamefont {F.}~\bibnamefont
  {Legoll}}\ and\ \bibinfo {author} {\bibfnamefont {T.}~\bibnamefont
  {Lelievre}},\ }\bibfield  {title} {\enquote {\bibinfo {title} {Effective
  dynamics using conditional expectations},}\ }\href@noop {} {\bibfield
  {journal} {\bibinfo  {journal} {Nonlinearity}\ }\textbf {\bibinfo {volume}
  {23}},\ \bibinfo {pages} {2131} (\bibinfo {year} {2010})}\BibitemShut
  {NoStop}%
\bibitem [{\citenamefont {Bhatt}\ and\ \citenamefont
  {Zuckerman}(2011)}]{bhatt2011beyond}%
  \BibitemOpen
  \bibfield  {author} {\bibinfo {author} {\bibfnamefont {D.}~\bibnamefont
  {Bhatt}}\ and\ \bibinfo {author} {\bibfnamefont {D.~M.}\ \bibnamefont
  {Zuckerman}},\ }\bibfield  {title} {\enquote {\bibinfo {title} {Beyond
  microscopic reversibility: Are observable nonequilibrium processes precisely
  reversible?}}\ }\href@noop {} {\bibfield  {journal} {\bibinfo  {journal}
  {Journal of Chemical Theory and Computation}\ }\textbf {\bibinfo {volume}
  {7}},\ \bibinfo {pages} {2520--2527} (\bibinfo {year} {2011})}\BibitemShut
  {NoStop}%
\bibitem [{\citenamefont {Best}\ and\ \citenamefont
  {Hummer}(2005)}]{best2005reaction}%
  \BibitemOpen
  \bibfield  {author} {\bibinfo {author} {\bibfnamefont {R.~B.}\ \bibnamefont
  {Best}}\ and\ \bibinfo {author} {\bibfnamefont {G.}~\bibnamefont {Hummer}},\
  }\bibfield  {title} {\enquote {\bibinfo {title} {Reaction coordinates and
  rates from transition paths},}\ }\href@noop {} {\bibfield  {journal}
  {\bibinfo  {journal} {Proceedings of the National Academy of Sciences}\
  }\textbf {\bibinfo {volume} {102}},\ \bibinfo {pages} {6732--6737} (\bibinfo
  {year} {2005})}\BibitemShut {NoStop}%
\bibitem [{\citenamefont {Rohrdanz}, \citenamefont {Zheng},\ and\ \citenamefont
  {Clementi}(2013)}]{rohrdanz2013discovering}%
  \BibitemOpen
  \bibfield  {author} {\bibinfo {author} {\bibfnamefont {M.~A.}\ \bibnamefont
  {Rohrdanz}}, \bibinfo {author} {\bibfnamefont {W.}~\bibnamefont {Zheng}}, \
  and\ \bibinfo {author} {\bibfnamefont {C.}~\bibnamefont {Clementi}},\
  }\bibfield  {title} {\enquote {\bibinfo {title} {Discovering mountain passes
  via torchlight: Methods for the definition of reaction coordinates and
  pathways in complex macromolecular reactions},}\ }\href@noop {} {\bibfield
  {journal} {\bibinfo  {journal} {Annual Review of Physical Chemistry}\
  }\textbf {\bibinfo {volume} {64}},\ \bibinfo {pages} {295--316} (\bibinfo
  {year} {2013})}\BibitemShut {NoStop}%
\bibitem [{\citenamefont {McGibbon}, \citenamefont {Husic},\ and\ \citenamefont
  {Pande}(2017)}]{mcgibbon2017identification}%
  \BibitemOpen
  \bibfield  {author} {\bibinfo {author} {\bibfnamefont {R.~T.}\ \bibnamefont
  {McGibbon}}, \bibinfo {author} {\bibfnamefont {B.~E.}\ \bibnamefont {Husic}},
  \ and\ \bibinfo {author} {\bibfnamefont {V.~S.}\ \bibnamefont {Pande}},\
  }\bibfield  {title} {\enquote {\bibinfo {title} {Identification of simple
  reaction coordinates from complex dynamics},}\ }\href@noop {} {\bibfield
  {journal} {\bibinfo  {journal} {The Journal of Chemical Physics}\ }\textbf
  {\bibinfo {volume} {146}},\ \bibinfo {pages} {044109} (\bibinfo {year}
  {2017})}\BibitemShut {NoStop}%
\bibitem [{\citenamefont {Suarez}, \citenamefont {Adelman},\ and\ \citenamefont
  {Zuckerman}(2016)}]{suarez2016accurate}%
  \BibitemOpen
  \bibfield  {author} {\bibinfo {author} {\bibfnamefont {E.}~\bibnamefont
  {Suarez}}, \bibinfo {author} {\bibfnamefont {J.~L.}\ \bibnamefont {Adelman}},
  \ and\ \bibinfo {author} {\bibfnamefont {D.~M.}\ \bibnamefont {Zuckerman}},\
  }\bibfield  {title} {\enquote {\bibinfo {title} {Accurate estimation of
  protein folding and unfolding times: beyond markov state models},}\
  }\href@noop {} {\bibfield  {journal} {\bibinfo  {journal} {Journal of
  Chemical Theory and Computation}\ }\textbf {\bibinfo {volume} {12}},\
  \bibinfo {pages} {3473--3481} (\bibinfo {year} {2016})}\BibitemShut {NoStop}%
\bibitem [{\citenamefont {Evans}(2010)}]{evans10}%
  \BibitemOpen
  \bibfield  {author} {\bibinfo {author} {\bibfnamefont {L.~C.}\ \bibnamefont
  {Evans}},\ }\href@noop {} {\emph {\bibinfo {title} {Partial differential
  equations}}}\ (\bibinfo  {publisher} {American Mathematical Society},\
  \bibinfo {address} {Providence, R.I.},\ \bibinfo {year} {2010})\BibitemShut
  {NoStop}%
\end{thebibliography}%

\end{document}